\newif\ifSC
\ifSC
\documentclass[onecolumn,draftclsnofoot,12pt]{IEEEtran}
\else
\documentclass[10pt,twocolumn]{IEEEtran}
\fi
\usepackage{siunitx}
\usepackage{gensymb}
\usepackage{amsthm}
\usepackage{mathrsfs}
\usepackage[dvipsnames]{xcolor}
\usepackage{booktabs,array}
\usepackage{tikz}
\usepackage{epsfig}
\usepackage{epstopdf}
\usepackage{graphicx}
\usepackage{amsfonts}
\usepackage{amssymb}
\usepackage{soul}
\usepackage{bbm}
\usepackage{url}

\usepackage{bm} 

\newcommand{\subsectionu}[1]{\subsubsection{#1}}



\usepackage{cite}
\DeclareMathAlphabet\mathbfcal{OMS}{cmsy}{b}{n}

\usepackage[cmex10]{amsmath}
\newif\ifmarkup
\markuptrue
\ifmarkup
  
  \newcommand{\akg}[1]{{\color{red}\textbf{[AKG:} #1\textbf{]}}}
  \newcommand{\chk}[1]{{\color{red}\textbf{[Check:} #1\textbf{]}}}
\else
  
  \newcommand{\akg}[1]{}
  \newcommand{\chk}[1]{}
\fi

\def\ie{{\em i.e.}~}

\usepackage{mathtools}
\newcommand{\lstate}[1]{\kappa_{#1}}
\newcommand{\app}[1]{%
\ifnum1=#1\relax
  \mathsf{M}
\else
 \mathsf{I}
\fi
}

\allowdisplaybreaks

\newcommand{\SINR}{\textrm{SINR}}

\newcommand{\laplace}{\mathcal{L}}

\newcommand{\bt}{\mathsf{b}}
\newcommand{\dv}{\mathrm{d}}

\newcommand{\ob}{\mathsf{o}}

\newcommand{\pt}{\mathrm{p}}
\newcommand{\rmL}{\mathrm{L}}
\newcommand{\rmN}{\mathrm{N}}

\newcommand{\x}{\mathbf{x}}
\newcommand{\z}{\mathbf{z}}

\newcommand{\crm}{\mathrm{c}}

\newcommand{\brm}{{\mathrm{b}}}
\newcommand{\urm}{{\mathrm{u}}}
\newcommand{\Grm}{{\mathrm{G}}}
\newcommand{\mrm}{{\mathrm{m}}}
\newcommand{\srm}{{\mathrm{s}}}
\newcommand{\arm}{{\mathrm{a}}}
\newcommand{\trm}{{\mathrm{t}}}
\newcommand{\rrm}{{\mathrm{r}}}
\newcommand{\qrm}{{\mathrm{q}}}
\newcommand{\ue}{\mathrm{u}}
\newcommand{\ueb}{\mathbf{u}}
\newcommand{\losprob}[2]{p_{#2}(#1)}

\newcommand{\pathlossBStoSEs}[1]{\ell_{#1,\srm}}

\newcommand{\pathlossSEtoUs}[1]{\ell_{#1,\ue}}

\newcommand{\pathlossBStoUs}[1]{\ell_{#1,\ue}}
\newcommand{\PDFY}{v}

\newcommand{\apIID}{\mathsf{I}}
\newcommand{\apMean}{\mathsf{M}}

\newcommand{\inserttable}{
	\begin{table}[t!]
		\centering
		\caption{Key parameters and their {default} values.}\label{tab:1}
		\resizebox{\columnwidth}{!}{			\begin{tabular}{|p{6.2cm}|p{2.8cm}|}
				\hline
				\textbf{Parameters}                        & \textbf{Values}                                           \\ \hline
				Carrier frequency ${f_{\mathrm{c}}}$ and {total} bandwidth ${W}$            & 0.3  \si{THz}, {500 \si{MHz}}                        \\ \hline
				Molecular absorption coefficient $\kappa_{f}$ & 0.001   \si{m^{-1}}                                      \\ \hline
				BS density  $\lambda_{\brm}$& {$10^{-4}$ \si{BS/m^2} }   \\ \hline
				{Blockage parameter $\beta$}          & {$0.02$ \si{m^{-1}}}         \\ \hline
				Main lobe gain of BS and UE $\Grm_{\mrm,\trm}=\Grm_{\mrm,\rrm}$  & 25 \si{dB}                                                \\ \hline
				Side lobe gain of BS and UE $\Grm_{\srm,\trm}=\Grm_{\srm,\rrm}$ & 0   \si{dB}                                                \\ \hline
				Scatterer density  $\lambda_{\srm}$          & 1 {\si{SE/m^2}}                                       \\ \hline
				Scattering {coefficient} $S$                 & 0.5 (rough surface)                                       \\ \hline
				Main lobe beamwidth {$\gamma_{\qrm}$} of BS and UE & 30\degree                                                 \\ \hline
				Thermal noise $N_{\ob}$    for $W=500$ \si{MHz}                & {$-77$ \si{dBm} }                 \\ \hline
				{UE density $\lambda_{\urm}$} & {$10^{-3}$ \si{UE/m^2}}                                    \\ \hline
				Pathloss {exponent} of LOS and NLOS $\alpha_{\rmL},\,\alpha_{\rmN}$ & {2 and 3{,} respectively}                                      \\ \hline
				{NLOS reflection loss $\xi$}          & {$0.1$ ($-10$ \si{dB})}                                \\ \hline
				{Scattering profile $\sprof{\cdot}$}   & {$\sprof{\cos\psi_{\x}}=\cos\psi_{\x}/2$}                          \\ \hline
				Effective SE area $A_{\srm}$        & {$10^{-2}$ \si{m^{2}}}       \\ \hline
				{SINR threshold $\tau$}               & {$20$ \si{dB}}                                        \\ \hline
				Transmit power $P_{\trm}$                  & {30 \si{dBm}}                                             \\ \hline
			\end{tabular}%

		}
		\par\vspace{2pt}\raggedright\footnotesize
	\end{table}
}

\newcommand{\R}{\mathbb{R}}

\newcommand{\pcov}{\mathrm{p_{c}}}

\usepackage{bbm}
\usepackage{amssymb}
\usepackage{amsmath,amsthm}
\usepackage{color}
\usepackage{amsfonts}
\usepackage{graphicx}
\usepackage{verbatim}
\usepackage{epstopdf}
\usepackage{cite}
\usepackage{tabulary}
\usepackage{mathtools}
\usepackage{enumitem}

\newcommand{\dd}{\mathrm{d}}

\newcommand\expect[1]{\mathbb{E}\left[#1\right]}
\newcommand\prob[1]{\mathbb{P}\left[#1\right]}

\newcommand{\expects}[2]{\mathbb{E}_{#1}\left[#2\right] }

\newcommand{\expU}[1]{e^{#1}}

\newcommand{\expS}[1]{\exp{\left(#1\right)}}

\def\home{\hbox{\kern3pt \vbox to13pt{}%
   \pdfliteral{q 0 0 m 0 5 l 5 10 l 10 5 l 10 0 l 7 0 l 7 5 l 3 5 l 3 0 l f
               1 j 1 J -2 5 m 5 12 l 12 5 l S Q }%
   \kern 13pt}}

\newcommand{\sprof}[1]{h(#1)}
\newtheorem{proposition}{Proposition}
\newtheorem{theorem}{Theorem}
\newtheorem{lemma}{Lemma}

\newtheorem{corollary}{Corollary}[theorem]

\allowdisplaybreaks
\begin{document}
	\title{
	Coverage of THz Networks in  Presence of Blockages and Scattering}
\author{Kaushlendra Pandey, Aman Pandey, Afaq A. Lone, Harpreet S. Dhillon, Abhishek K. Gupta \,\,\vspace{-2.1em}
	\thanks{K. Pandey is with CIT Kokrajhar, India (Email: kk.pandey@cit.ac.in). A. Pandey, A. A. Lone and A. K. Gupta are with IIT Kanpur, India 208016 (Email: \{amanpandey20,afaq,gkrabhi\}@iitk.ac.in). H. S. Dhillon is with Wireless@VT, Department of ECE, Virginia Tech, Blacksburg, VA, USA (Email: hdhillon@vt.edu). The research is supported by Qualcomm 6G research grant. A part of this paper was presented in ICC 2023 \cite{pandey2023coverage}.}
}
	\maketitle
	\begin{abstract}
In this paper, we develop an analytical framework to study a cellular THz network that incorporates the effects of blockages, scattering elements (SEs), directionality, atmospheric absorption and noise. We first characterize the power incident on each SE and the power scattered towards the users, while considering the presence of blockages. Modeling the SEs via a Poisson point process (PPP), we derive the Laplace transform (LT) of the sum interference caused by the SEs at the typical user. We also characterize the sum interference caused by the interfering base stations (BSs). Using this framework, we derive the coverage probability (CP), the mean signal to interference plus noise ratio (SINR), the rate coverage (RC) and the area spectral efficiency (ASE) at the typical user, and we validate them against a Monte Carlo simulation that retains the correlation of the incident power across SEs. The analysis reduces the design of the network to a few quantities that follow from the system parameters. We also identify various densification regimes highlighting two {critical/corner values} of BS density and show that {the} shape of the coverage versus densification curve along with the optimal density value depends on these two critical densities. We identify a critical SE density, via equating the mean SE interference with the typical cell edge signal strength, which marks the deployments in which scattering must be accounted for. Below this density, the SE interference has a small effect on the coverage, whereas above it, the coverage falls steeply. We show that the critical density doesn't depend signficantly on the BS density. Result shows that, in heavily cluttered deployments with significant number of SEs, their interference must be accounted for to get accurate results.
	\end{abstract}
	\section{Introduction}
	{The 5G networks have explored the mmWave bands up to 70 GHz to support high data rate communication owing to the availability of large mmWave spectrum and highly directional communication. Despite this, the emergence of applications such as virtual and augmented reality, ultra HD video conferencing, 3D gaming and medical devices \cite{tripathi2021millimeter} brings forth more stringent requirements for future 6G networks. The THz band in particular the 100--300 \si{GHz} spectrum is a promising candidate for future networks and can provide data rates of the order of {Tbps}.  Despite the potential, communication in THz bands suffers from various challenges beyond those faced by mmWave due to small antenna sizes, highly directional communication, and high sensitivity to blockage. First, {THz communication} suffers from atmospheric absorption which reduces the signal power exponentially with distance \cite{tripathi2021millimeter}. Further, the small wavelength at THz frequencies results in a significant scattering compared to lower frequencies \cite{tripathi2021millimeter}, resulting in an additional interference. These factors can significantly degrade the performance of THz networks. Therefore, a comprehensive analysis of THz network considering all the factors impacting the THz propagation is crucial to understand the potential of THz communication for future applications.}

\subsubsection{Related work}
The prior work has studied the performance of THz communication at the link level. The use-cases of the $100$ \si{GHz}--$3$ \si{THz} band for 6G are surveyed in
\cite{rappaport2019access,tripathi2021millimeter}. {More recent overviews of the physical-, link- and network-layer challenges of THz systems are given in \cite{sarieddeen2021overview,chaccour2022seven,shafie2023terahertz,jiang2024terahertz,jornet2024evolution}, and the performance and implementation road-map of last-meter indoor THz access is discussed in \cite{petrov2018last}.} The paper \cite{hirata2015ultrafast} provides an extensive overview of the development and implementation of THz networks. Due to the small aperture of THz antennas, large arrays are required to satisfy the link budget. The array-of-subarrays architecture and associated hybrid beamforming designs are discussed in \cite{lin2016array}. The challenges in THz transceivers and massive phased arrays are reviewed in \cite{Challenges_THz}. From the perspective of channel modeling, \cite{moleculecoeff} characterizes various propagation losses including the
	molecular absorption loss and path loss which limit the coverage distance  at THz \cite{akyildiz2018combating}. The works \cite{han2015multiray,priebe2013stochastic} develop multi-ray and
	stochastic THz channel models. {A comprehensive survey of THz channel measurements and models, covering the deterministic, stochastic and hybrid approaches and the treatment of rough-surface scattering, is given in \cite{han2022terahertz}.} Scattering is a distinguishing feature of THz bands as compared to mmWave \cite{tripathi2021millimeter,degli2007measurement,ju2019scattering}. At THz wavelengths, usually smooth surfaces such as ordinary building surfaces start appearing rough, resulting in the scattering of a significant part of incident power away from the specular direction \cite{tripathi2021millimeter}. While these above-mentioned works focus on link-level performance, effect of network geometry, blockages, scattering and interference can only be understood from a  system-level analysis. {The stochastic geometry (SG) has emerged as a foundational approach to derive system level analysis of wireless networks due to its ability to model various geometry related factors \cite{andrews2023introduction}. In particular, it can be used to model blockages via Boolean process \cite{andrews2016modeling}, locations of BSs and users (UEs) \cite{andrews2011tractable} via point processes	and directionality \cite{andrews2023introduction}. {A tutorial on the mathematical modeling of mmWave/THz cellular systems, covering directional antennas, dynamic blockage and user micro-mobility, is given in \cite{andrews2016modeling,moltchanov2022tutorial}.}} {The authors in \cite{petrov2015interference} described the distribution of interference and SINR in a dense THz network while considering the effect of antenna directivity and molecular absorption. The outage probability was calculated for a THz cellular network in \cite{kokkoniemi2017stochastic}. However, these papers have not incorporated the effect of blockages or the influence of scattering. The effect of blockage was studied in \cite{petrov2017interference} using SG showing that the antenna directivity and blockages can help in reduction of mean interference at a user. However, in this work, authors have not considered effect of non-line-of-sight (NLOS) propagation and scattered interference. \cite{yao2017stochastic} incorporates the effect of NLOS propagation along with line-of-sight (LOS) propagation, antenna directivity and blockages, without considering the effect of scattering. Similarly, in \cite{moltchanov2018analytical}, the authors used Talbot’s technique to invert the Laplace transform (LT) of the interference and to get corresponding probability density functions (PDFs). The coverage of dense THz networks, 3D THz networks, coexisting RF/THz networks and multi-tier THz networks were studied in \cite{sayehvand2020interference,raja2021coverage,Shafie_2021} and  \cite{kouzayha2022coexisting}.} \cite{wu2021interference} analyzed the coverage of an indoor THz network with wall and human-body blockages under nearest-LOS access-point association, {while} \cite{tang2025coverage} extends this to a 3D indoor layout with multi-cluster fluctuating two-ray fading{.} The ergodic outage and capacity of a THz link under micro-mobility and blockage are characterized in \cite{moltchanov2022ergodic}, but at the link rather than the network level. {Further, all of these works do not consider the impact of scattering.} Moreover, the significance of SE interference relative to noise and BS interference, its dependence on BS densification, and its validation under correlated scattering remain unexplored. Despite recent works on THz networks, there is lack of a comprehensive framework that incorporates the impact of {scatterers}, blockages, directionality, and molecular absorption hindering the understanding of interplay of these essential parameters in the performance of THz networks, which is the goal of this paper. Table~\ref{tab:comparison} {provides a summary of the contributions of} this work against the existing literature.

\newcommand{\adj}{\hspace{-.08in}}

	\begin{table}[t!]
		\caption{Comparison of this work with the existing stochastic-geometry based analyses of THz networks.}\label{tab:comparison}
\footnotesize
		\begin{tabular}{p{.12in}p{.2in} p{.15in}p{.18in}p{.18in}p{.2in}p{.24in}p{.9in}}
			\hline
			\hspace{-.03in}\textbf{Wo rk} & \textbf{LOS/ NLOS links}
			& \textbf{Bloc kag es}
			& \textbf{Mol. abso rption}
			& \textbf{Ant. side lobes}
			& \textbf{Scatte ring elements}
			& \textbf{Densifi cation analysis}
			& \textbf{Performance metrics}\\
			\hline\adj
\cite{petrov2015interference}& $\times$     & $\times$     & $\checkmark$ & $\times$     & $\times$          & $\times$     & Interference, SINR \\\hline
				\adj	\cite{kokkoniemi2017stochastic}   & $\times$     & $\times$     & $\checkmark$ & $\times$     & $\times$          & $\times$     & Mean interference, outage \\
			\adj\cite{petrov2017interference}     & $\times$     & $\checkmark$ & $\checkmark$ & $\times$     & $\times$          & $\times$     & Mean interference, SINR \\
	
			\adj\cite{yao2017stochastic}          & $\checkmark$ & $\checkmark$ & $\checkmark$ & $\times$     & $\times$          & $\times$     & CP \\
			\adj\cite{moltchanov2018analytical}   & $\checkmark$ & $\checkmark$ & $\checkmark$ & $\times$     & $\times$          & $\times$     & Interference/signal-to-interference ratio (SIR) densities \\
			\adj\cite{sayehvand2020interference}  & $\checkmark$ & $\checkmark$ & $\checkmark$ & $\times$     & $\times$          & $\times$     & CP, rate CP \\
			\adj\cite{raja2021coverage}           & $\checkmark$ & $\checkmark$ & $\checkmark$ & $\times$     & $\times$          & $\times$     & CP (multi-tier) \\
			\adj\cite{Shafie_2021}                & $\checkmark$ & $\checkmark$ & $\checkmark$ & $\times$     & $\times$          & $\times$     & CP (3D) \\
			{\adj\cite{kouzayha2022coexisting}} & {$\checkmark$} & {$\checkmark$} & {$\checkmark$} & {$\times$} & {$\times$} & {$\times$} & {CP, rate} \\
			\adj{\cite{wu2021interference}}    & {$\checkmark$} & {$\checkmark$} & {$\checkmark$} & {$\times$} & {$\times$} & {$\checkmark$} & {Mean interference, CP, throughput} \\
			\adj{\cite{tang2025coverage}}      & {$\checkmark$} & {$\checkmark$} & {$\checkmark$} & {$\times$} & {$\times$} & {$\checkmark$} & {CP (3D indoor)} \\
\hline
			\hspace{-.1in}\textbf{This work}                & $\checkmark$ & $\checkmark$ & $\checkmark$ & $\checkmark$ & $\checkmark$  & $\checkmark$ & CP, mean SINR, rate CP, ASE \\
			\hline
		\end{tabular}\vspace{-.23in}
\end{table}

\subsubsection{Contributions} 
In this paper, we develop an analytical framework to study a cellular THz network that incorporates blockages, SEs, directionality, atmospheric absorption, noise, and BS densification. The key contributions {of the work are} as follows.
{
\begin{enumerate}
	\item We develop an analytical framework for a cellular THz network {which incorporates BSs (modeled as a PPP), their interference, scatterer interference (with scatterers modeled as a PPP), and the effect of blockages (modeled as a Boolean process). We} include directional antennas with side-lobe radiation at both the BSs and the UEs.
	\item We {present an alternative derivation of} the THz path-loss due to atmospheric absorption from a Boolean--Poisson model of the absorbing molecules.
	\item We characterize the power incident on a SE and the power it scatters towards the UE under blockage. We derive the LT of the SE interference under two tractable approximations of the correlated incident power, the mean and the {independent and identically distributed (IID)} approximation.
	\item We characterize the interference from the interfering BSs and derive the CP, the rate coverage (RC) and the area spectral efficiency, and we obtain the mean SINR through an auxiliary-variable argument.
	\item {Utilizing a} correlation-preserving Monte Carlo simulation{, we validate the analysis and derive various insights. In particular, we identify various densification regimes. We highlight two critical/corner densities of BSs and show that the shape of the coverage versus densification curve along with the optimal density depends on these two critical densities. We also show the presence of} {a critical SE density beyond which scattering must be accounted for, and show that it is nearly independent of the BS density.}
\end{enumerate}}
\noindent\textit{Notation:} The notation $\x$ stands for a $2$D vector with its norm $\|\x\|=x$. $\bt(\x,r)$ denotes a ball with radius  $r$ and center  $\x$. For random variable (RV) $X$, its probability generating function (PGF), cumulative distribution function (CDF), and PDF are written as $\mathcal{P}_{X}(s),$ $F_{X}(x),$ and $f_{X}(x)$, respectively.  $c$ denotes the speed of light. For a non negative RV $X${, $\mathbb{E}[X]=\int_{0}^{\infty}\prob{X>\tau}\dv \tau$.}
The key notation is summarized  in Table~\ref{tab:notation}. The following {two} mathematical results are used throughout this paper {(see Appendix~A of the supplementary material \cite{pandey2026supplementary} for proofs)}.
\begin{proposition}\label{proposition}
	{For two functions $f(\cdot)$ and $b(\cdot)$, we have}
	\begin{align}
		\int\nolimits_{0}^{\pi} f\left(b(x) |\sin(\omega-\Omega)|\right)\dv \omega=\int\nolimits_{0}^{\pi}f\left(b(x)\sin t\right)\dv t{.} \label{eq:prop1}
	\end{align}
\end{proposition}
\begin{proposition}\label{propos2}
	If $\omega$ is a uniformly distributed RV between $0$ and $\pi$ and $Y=|\sin (\Omega-\omega)|${, the} PDF of $Y$ is {$\PDFY(y)={2}/\big({\pi\sqrt{1-y^2}}\big)$ for $0\leq y\leq 1$.}
\end{proposition}

\begin{table}[t!]
	\caption{Key notation}\label{tab:notation}
	\centering\footnotesize
	\setlength{\tabcolsep}{2.4pt}
	\renewcommand{\arraystretch}{1}
\begin{tabular}{p{.12\textwidth}p{.36\textwidth}}
		\hline
		\textbf{Symbol} & \textbf{Description} 
\\
		\hline
		$\Phi_{\brm}$, $\lambda_{\brm}$ & PPP of BSs and its density 
		\\
		 $P_{\trm}$, $N_{\ob}$ & Transmit power, noise power \\
		$\Phi_{\lstate{},\brm}${, $\Phi_{\lstate{},\srm}$} & PPP of LOS ($\lstate{}=\rmL$) and NLOS ($\lstate{}=\rmN$) BSs {and SEs} 
		\\
		 $\kappa_f$ & Molecular absorption coefficient \\
		$\Phi_{\srm}$, $\lambda_{\srm}$ & PPP of SEs and its density \\
		 $\alpha_{\rmL}$, $\alpha_{\rmN}$ & LOS and NLOS pathloss exponents \\
		$\z$, $\x$, $\ueb$ & Location of a BS, of an SE, of the typical UE \\
		 $\pathlossBStoUs{\lstate{}}(\cdot)$ & Pathloss of a link ending at the UE \\
		$R_{\ob}$ & Serving distance (nearest LOS BS) \\
		 $\pathlossBStoSEs{\lstate{}}(\cdot)$ & Pathloss of a link ending at an SE \\
		$\beta$ & Blockage parameter \\
		 $\gamma_{\lstate{},\arm}$ & Pathloss coefficient of a $\lstate{}$ link to device $\arm$ \\
		$\lstate{l}$ & State of link $l$, $\lstate{l}\in\{\rmL,\rmN\}$ \\
		 $S$, $\sprof{\cdot}$ & Scattering coefficient, scattering profile \\
		$g_{\qrm}(\theta)$ & Antenna pattern of device $\qrm\in\{\trm,\rrm\}$ \\
		 $\psi_{\x}$ & Scattering angle at the SE at $\x$ towards the UE \\
		$\Grm_{\mrm,\qrm}$, $\Grm_{\srm,\qrm}$, $\gamma_{\qrm}$ & Main-lobe gain, side-lobe gain, beamwidth \\
		 $\Omega_{\x}$, $\omega_{\x}$ & Angular location and orientation of the SE at $\x$ \\
		$\pt_{\trm}$, $\pt_{\rrm}$ & Main-lobe probability at the BS, at the UE \\
		 $\theta_{\z_i}$, $\delta_{\z_i}$ & Incidence angle and direction of BS $\z_i$ at the SE \\
		$\Grm_{i}$, $\pt_{i}$ & BS--UE gain pair $i$ and its probability, \eqref{eq:AntennaGainValues} \\
		 $\widetilde{\beta}_{\z\x}$, $\widetilde{\beta}_{\x}$ & BS beam angle to  SE, UE beam angle to  SE \\
		$\Grm_{\qrm,i}$, $\pt_{\qrm,i}$ & Lobe gain of device $\qrm$ and its probability, $i=1,2$ \\
		 $\phi_{\z_i}$ & Beam orientation of BS $\z_i$ \\
		$P_{\x}$, $P_{\x_{\ob}}$ & Power incident on the SE at $\x$, on the typical SE \\
		 $\mathrm{U}_{\z\x\ueb}$ & BS at $\z$ and UE on the same side of the SE at $\x$ \\
		$m_{\ob}$ & Mean incident power on the typical SE, $\mathbb{E}[P_{\x_{\ob}}]$ \\
		 $\mathcal{S}(R_{\ob})$ & Received power from the serving BS \\
		$I_{\brm}$, $I_{\srm}$ & Interference from BSs, from SEs \\
		 $\tau$, $\rho$ & SINR threshold, rate threshold \\
		$I_{\srm,\apMean}$, $I_{\srm,\apIID}$ & $I_{\srm}$ under the mean / IID $P_{\x}$ approximation \\
		 $\mathrm{p}_{\mathrm{c}}$, $\tilde{\mathrm{p}}_{\mathrm{c}}$, $\mathrm{p}_{\mathrm{R}}$ & CP, complementary CDF (CCDF) of the signal to SE interference plus noise ratio (SSIR), RC \\
		$\laplace_{X}(s)$ & LT of $X$ \\
		 $\overline{\mathcal{R}}$, $\mathcal{A}$ & Average ergodic rate, area spectral efficiency \\
		\hline
	\end{tabular}\vspace{-.2in}
\end{table}
\section{System model}
In this paper, we consider a THz cellular network {consisting} of multiple THz BSs and UEs in the presence of blockages and SEs. The location of THz BSs is modeled using $2$D PPP $\Phi_{\brm}=\{ \z_{i},\,\forall i \in \mathbb{N} \}$ with density $\lambda_{\brm}$. We assume that each BS transmits {with} the power $P_\trm$ to {communicate} with UEs. The locations of UEs {are} modeled with an independent and homogeneous PPP $\Phi_\urm$ which has density $\lambda_\urm$. UEs and BSs are outfitted with highly directional antennas.
\subsubsection{Antenna Model}\label{sec:antenna}
We consider the widely used cone model \cite{di2015stochastic} to describe the antenna{'s} directivity where the antenna gain in a desired direction is given as  
\begin{align}\label{Gain}
	g_{\qrm}(\theta) =\begin{cases} \Grm_{\mrm,\qrm},&  \vert{\theta}\vert\leq \gamma_{\qrm}/2\\
		\Grm_{\srm,\qrm},& \text{else},
	\end{cases}
\end{align}where $\qrm \in \{\trm,\rrm\}$ denotes the device type \ie transmitter or receiver device. Here $\Grm_{\mrm,\trm},\Grm_{\srm,\trm},\Grm_{\mrm,\rrm},\Grm_{\srm,\rrm}$ denote the main and the side lobe gains for transmit and receive  antenna respectively. Further $\theta\in [-\pi,\,\pi)$ denotes the angle of desired direction in relation to beam orientation of the antenna \ie the axis of the maximum gain and  $\gamma_{\qrm}$ is the antenna beam width. Assuming ${N_{\qrm}}$ elements at the antenna, we can take $\Grm_{\mrm,\qrm}=N_{\qrm}$, the major lobe beamwidth as $\gamma_{\qrm}$ = $\frac{\Omega_{0}}{N_{\qrm}}$ \si{degrees} and minor lobe beamwidth as $\tilde{\gamma}_{\qrm}=360^{\circ}-\frac{\Omega_{0}}{N_{\qrm}}${, where $\Omega_{0}$ is a constant \cite{GupAlkAndHea2016}. Further,} to normalize the transmitted power, we have
${(\Grm_{\mrm,\qrm}\gamma_{\qrm}+\Grm_{\srm,\qrm}\tilde{\gamma}_{\qrm})}/{(2\pi)}{=}1{,}$ which is due to the fact that the power  directed from BSs/UE antenna in all directions does not exceed the total transmit power\cite{sayehvand2020interference,di2015stochastic}. Note that the gain and beamwidth values in Table~\ref{tab:1} describe a beam that is narrow in both azimuth and elevation directions. Using beam alignment methods, the typical UE and its associated BS align themselves so that their main lobes coincide. However, an interfering BS does not need to align with the typical UE{. Therefore, for an interfering link,} the directional gain $g_{\z}$ is a RV with the distribution given as
\begin{align}
\!\!\!\! g_{\z}\!\!=	\!\!\begin{cases} \Grm_{1}\!=\!\Grm_{\mrm,\trm}\Grm_{\mrm,\rrm},& \!\!\! \!\text{with prob. $\pt_{1}\!=\!\pt_{\trm}\pt_{\rrm}$ }\\
		\Grm_{2}\!=\!\Grm_{\mrm,\trm}\Grm_{\srm,\rrm},& \!\!\!\!\text{with prob.  $\pt_{2}\!=\!\pt_{\trm}(1-\pt_{\rrm})$ } \\
		\Grm_{3}\!=\!\Grm_{\srm,\trm}\Grm_{\mrm,\rrm}, &\!\!\!\!\text{with prob. $\pt_{3}\!=\!(1-\pt_{\trm})\pt_{\rrm}$ }\\
		\Grm_{4}\!=\!\Grm_{\srm,\trm}\Grm_{\srm,\rrm}, &\!\!\!\!\text{with prob. $\pt_{4}\!=\!(1-\pt_{\trm})(1-\pt_{\rrm}),$ }
	\end{cases}
	\label{eq:AntennaGainValues}
\end{align}
where $\pt_{\rrm}={\gamma_{\rrm}}/(2\pi)$ denotes the probability of receiving maximum gain at the UE antenna. Similarly, $\pt_{\trm}={\gamma_{\trm}}/(2\pi)$ is defined for the  BS antenna.

		\subsubsection{Modeling of Blockages} {We consider that various buildings and obstacles can block the signal. At THz frequencies, the blockage can be significant and result in the total outage of links. The buildings can be modeled using a Boolean model {with their} centers distributed as PPP and {orientations distributed uniformly}. Let $\losprob{z}{\rmL}$ denote the probability that a link of length $z$ is LOS. Under the Boolean model, we have $\losprob{z}{\rmL}= e^{-\beta z}$ where $\beta$ is a constant that depends on characteristics of the blockages such as blockages' density $\mu$ and size. For example, when buildings are considered as line segments of mean length $L$, then $\beta= (2/\pi) \mu L$. From the typical user perspective, BSs can be categorized into LOS and NLOS BSs. {Assuming independent blockages, from} the thinning theorem \cite{andrews2023introduction}, the locations of LOS BSs follow a PPP, denoted as $\Phi_{\rmL,\brm}$, with density $\lambda_{\rmL,\brm}(z) = \lambda_{\brm} e^{-\beta z}$. Similarly, the locations of NLOS BSs form a PPP $\Phi_{\rmN,\brm}$ with density $\lambda_{\rmN,\brm}(z) = \lambda_{\brm} (1 - e^{-\beta z})$.} Let $\losprob{x}{\rmL}=e^{-\beta x}$ and $\losprob{x}{\rmN} =1-e^{-\beta x}$ denote the LOS and NLOS probabilities.

 \subsubsection{Modeling of {Scatterers}}
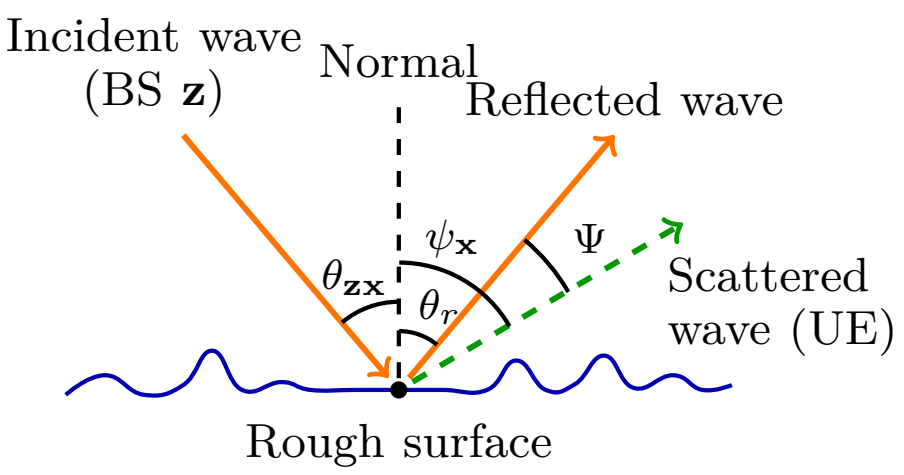
\begin{figure}[t!]
	\centering
	\begin{tikzpicture}[x=0.72cm,y=0.72cm,font=\footnotesize,scale=.9]
		\draw[blue!70!black, thick] plot[smooth, tension=0.8] coordinates
		{(-3.4,-0.05) (-3.0,0.15) (-2.6,-0.05) (-2.2,0.05) (-1.9,0.40) (-1.6,0.0)
		 (-1.2,0.08) (-0.9,0.0) (-0.45,0.0) (0,0) (0.45,0.0) (0.9,0.0) (1.2,0.30)
		 (1.5,0.0) (1.8,0.06) (2.1,0.35) (2.4,0.02) (2.8,0.15) (3.1,0.0) (3.4,0.05)};
		\node[below, font=\small\bfseries] at (0,-0.15) {Rough surface};
		\draw[dashed, thick] (0,0.12) -- (0,3.0);
		\node[above, font=\small\bfseries] at (0,3.0) {Normal};
		\draw[orange!90!red, line width=1.1pt, ->, shorten <=3pt] (0,0) -- (2.2,2.6);
		\node[above, font=\small\bfseries] at (2.3,2.6) {Reflected wave};
		\draw[green!60!black, line width=1.1pt, dashed, ->, shorten <=3pt] (0,0) -- (2.9,1.7);
		\node[below right, align=left, font=\small\bfseries] at (2.55,1.55) {Scattered\\wave (UE)};
		\draw[orange!90!red, line width=1.1pt, ->, shorten >=3pt] (-2.2,2.6) -- (0,0);
		\node[above, align=center, font=\small\bfseries] at (-2.5,2.6) {Incident wave\\(BS $\z$)};
		\fill (0,0) circle (1.8pt);
		\draw[thick] (0,0.9)     arc (90:130:0.9);  \node at (-0.45,1.15) {$\theta_{\z\x}$};
		\draw[thick] (0.39,0.46) arc (50:90:0.6);   \node at (0.42,0.85)  {$\theta_{r}$};
		\draw[thick] (1.13,0.65) arc (30:90:1.3);   \node at (0.55,1.55)  {$\psi_{\x}$};
		\draw[thick] (1.73,1.0)  arc (30:50:2.0);   \node at (1.95,1.55)  {$\Psi$};
	\end{tikzpicture}
	\caption{Scattering from a rough surface (SE at $\x$), {where} $\theta_{\z\x}$ is the angle of incidence of the wave from the BS at $\z$, $\theta_{r}$ the specular reflection angle, $\psi_{\x}$ the scattering angle towards the UE, and $\Psi$ the angle between the reflected and the scattered waves.}\vspace{-.15in}
	\label{fig:scattering}
\end{figure}

 Due to the smaller wavelength at THz frequencies, otherwise smooth surfaces can appear rough, and result in significant scattering compared to lower frequencies \cite{tripathi2021millimeter}, \cite{han2022terahertz}. Scattering can introduce additional interference, thereby degrading system performance. Therefore, this work {incorporates} the impact of scattering on system performance. We consider that the environment is filled with {many SEs} distributed everywhere, {and} hence their locations can be modeled as PPP $\Phi_{\srm}=\{\x_{i},\,\x_{i}\in \R^{2}\}$ with density $\lambda_\srm$. 
 Each SE scatters the power incident on its surface from {each} BS. The scattered power is distributed in all directions on the same side as the incident wave. We consider the {Lambertian} scattering model, which states that the scattered power in any direction is proportional to the cosine of the scattering angle {$\psi$} \cite{tripathi2021millimeter}{. In general,}
 \begin{equation}
 	P_{\mathrm{S}}({\psi})
 	\propto S^{2}{\sprof{\cos({\psi})}},\label{eq:scatprofile}
 \end{equation}
 where \(S\) denotes the scattering coefficient, {with $S^{2}$} {representing} the fraction of the incident power that is scattered from the surface of the SE \cite{tripathi2021millimeter}{, and $\sprof{\cdot}$ is the scattering directivity function. Examples of $\sprof{\cdot}$ include {$\sprof{x}=x/2$, the normalized Lambertian model used in this paper, for which $\int_{-\pi/2}^{\pi/2}\sprof{\cos\psi}\dv \psi=1$ and hence the total scattered power equals the scattered fraction of the incident power.}} Physically, consider a wave emitted by {a} BS at $\z$ and incident on {a} SE at $\x$, at an angle $\theta_{\z\x}$ to the surface normal (Fig.~\ref{fig:scattering}). For a surface smooth on the wavelength scale, a fraction $\Gamma_{\srm}^{2}$ of the incident power is reflected. For a rough surface, the wave is scattered into multiple directions in addition to the specularly reflected component. By the Rayleigh criterion, a surface is rough when its height variations exceed $c/(8f_{\mathrm{c}}\cos\theta_{\z\x})$ \cite{tripathi2021millimeter,degli2007measurement}.{ At millimetre wavelengths, most building surfaces are rough, and the scattered fraction $S^{2}$ of the incident power increases with surface roughness, with $S^{2}+\Gamma_{\srm}^{2}\leq 1$ \cite{ju2019scattering}.}
\begin{figure}[t!]
	\centering
	\includegraphics[width=0.6\linewidth]{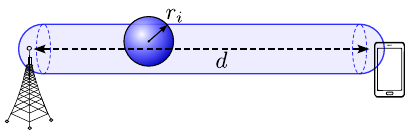}
\caption{Illustration of the region $\mathcal{B}_i\oplus K$ (shaded) in which the center of a molecule of radius $r_i$ must lie to intersect a link of length $d$.}\label{fig:molecularabsorption}\vspace{-.15in}
\end{figure}

\subsubsection{Pathloss Model}
	We {consider} the bounded pathloss model \cite{gupta2022impact} {with} the pathloss for NLOS and LOS links represented by $\ell_{\rmN,\arm}(y)$ and $\ell_{\rmL,\arm} (y)$,  respectively. {Here,} $\arm \in \{\ue,\srm\}$. We assume that the pathloss model satisfies the following realistic conditions
		\begin{enumerate}
			\item $\ell_{\rmL,\arm}(y)>\ell_{\rmN,\arm}(y),\forall y$.
			\item $\ell_{\rmL,\arm}(y_{1})>\ell_{\rmL,\arm}(y_{2})$ and $\ell_{\rmN,\arm}(y_{1})>\ell_{\rmN,\arm}(y_{2}),\, \forall y_{1}<y_{2}$.
			\item $C_{\rmL}\!=\!\max_{y}(\ell_{\rmL,\arm}(y))\!<\!\infty$\! and  $  C_{\rmN}\!=\!\max_{y}(\ell_{\rmN,\arm}(y))\!<\!\infty$.
		\end{enumerate}
The first condition suggests that a LOS link carries more power as compare{d} to {a} similar NLOS link of the same distance, the second condition says that the received power monotonically decreases with {the} link length, and the third condition suggests that the maximum power at the receiver is bounded. 
 \subsubsection{Cell Association Model}
 We consider the closest LOS BS association where each 
 UE connects to the closest BS among all LOS BSs. Such
 association model is assumed due to the considerable effects
 of blockages on THz waves and by using the fact that in case
 of dense THz networks {LOS links are} more dominant
 than NLOS \cite{sayehvand2020interference}. To characterize the network's performance, we consider the typical UE which can be taken at the origin without loss of generality. Let $R_\ob$ denote the distance of the serving BS to the UE. 
The serving signal strength at the UE  is given as 
\begin{align}
{\mathcal{S}}(R_{\ob})&={P_{\trm}}\Grm_{\mrm,\trm}\Grm_{\mrm,\rrm}\pathlossBStoUs{\rmL}(R_{\ob}).%
	\label{Sro}
\end{align} 
Hence, the $\SINR$ at the typical UE is provided as
\begin{align*}
	&\SINR=\frac{{\mathcal{S}}(R_{\ob})}{I_{\brm}+I_{\srm}+N_{\ob}},
	\end{align*}
where
$I_{\brm}$ is the interference due to BSs, $I_{\srm}$  is the SE interference and $N_\ob$ is the noise. We aim to compute the CP which is defined as the probability that the $\SINR$ at the typical UE is greater than a certain threshold $\tau$ \ie,
	${\rm p_{c}}(\tau)=\prob{\SINR>\tau}$.
To proceed further, we need to characterize the
interference from BSs and SEs via their LTs. 

\section{Impact of Atmospheric Absorption}
Let us consider a communication link of length \( d \), between a THz BS and a UE, as shown in Fig.~\ref{fig:molecularabsorption}. The surrounding environment contains molecules {of many types, with the molecules of each type $i$ distributed as} a homogeneous PPP with density \( \lambda_i \) and effective radius \( r_i \).
Thus, the spatial distribution of these molecules can be modeled as a {Boolean-Poisson process}. {Let} \( {\nu_i} \) {denote the number of} molecules of type \( i \) that intersect the communication link{. From random shape theory, the intersection creates an independent thinning of the PPP and therefore, from the independent thinning theorem, the intersecting molecules also form a PPP with their number $\nu_i$ distributed as a} Poisson {RV} with {mean}
\begin{align*}
	\mu_i = \lambda_i \, |{\mathcal{B}_i} \oplus K| = \lambda_i\left( ({4}/{3})\pi r_i^3 + \pi r_i^2 d \right),
\end{align*}
where \( {\mathcal{B}_i} \) denotes a sphere of radius \( r_i \), \( K \) denotes the line segment (link) of length \( d \) and \( \oplus \) denotes the Minkowski sum. For practical scenarios where \( r_i \ll d \), the mean can be approximated as
\(	\mu_i \approx \lambda_i \pi r_i^2 d = \lambda_i d A_{\text{eff},i},\)
where \( A_{\text{eff},i} = \pi r_i^2 \) denotes the effective cross-sectional area {of molecule type $i$ at the operating frequency}. {Considering that each intersection of the signal with a molecule results in an attenuation by a factor} \( \eta_i \), the expected signal attenuation is
\begin{align}
	H_i &= \mathbb{E}[\eta_i^{{\nu_i}}] = e^{-\mu_i}\textstyle\sum_{k\geq0} \frac{(\eta_i\mu_i)^k}{k!} = e^{-\mu_i (1 - \eta_i)} .
	\label{eq:Hi}
\end{align}
Hence, the effective received power due to molecular absorption is
\begin{align}
	P_{\rrm} &= P_{\trm} \prod\nolimits_{i} H_i = P_{\trm} \exp\Big( -\sum\nolimits_{i} \lambda_i d A_{\text{eff},i} (1 - \eta_i) \Big) \nonumber\\
	&= P_{\trm} \exp(-\kappa_f d),\quad
	\text{with } \kappa_f = \sum\nolimits_{i} \lambda_i A_{\text{eff},i} (1 - \eta_i)\label{eq:kf}
\end{align}{denoting the molecular absorption coefficient. It can be seen that $\kappa_f$ depends on the operating frequency, the composition of the environment and the relative cross-sections of its molecules.}

\noindent{Combining the effect of molecular absorption with the standard path-loss due to the spreading of the signal, the effective path-loss for a LOS and an NLOS link is given as}
$\ell_{\rmL,\arm}(z)=\min (1,\gamma_{\rmL,\arm} z^{-\alpha_{\rmL}}e^{-\kappa_{f}z}),\,\, \ell_{\rmN,\arm}(z)=\min (1, \gamma_{\rmN,\arm}z^{-\alpha_{\rmN}}e^{-\kappa_{f}z}),$	respectively. Here
$\alpha_{\rmL}$ and $\alpha_{\rmN}$ {denote the} pathloss exponent{s} for LOS and NLOS link{s}, respectively. For a link between BS-UE or SE-UE, the LOS coefficient is given as $\gamma_\mathrm{L,u} = \frac{A_e}{4\pi}$ where $A_e=\frac{c^2}{4\pi f_{\mathrm{c}}^2}$ denotes the effective antenna aperture.
For link between BS-SE, the LOS coefficient is given as $\gamma_\mathrm{L,s} = \frac{A_\srm}{4\pi}$ where $A_{\srm}$ denotes the effective SE area.
Further, as stated in \cite{sayehvand2020interference} LOS transmission is dominant compared to NLOS transmission. For NLOS links, $\gamma_{\rmN,\arm}=\xi \gamma_{\rmL,\arm}$, {with} $\xi$ {denoting the losses} due to NLOS reflections.
\section{Characterization of Interference Due to SEs}\label{sec:SEinterference} 
The computation of SE interference at the UE requires characterization of the power incident on {each} SE, scattering profile at SEs and the power received at the UE from these SEs, which we compute separately in the following sections.
\subsection{{Scattered Power for a Single BS--SE--UE Link}}
\begin{figure}[t!]
		\centering
\begin{minipage}{0.27\textwidth}
\includegraphics[width=\linewidth]{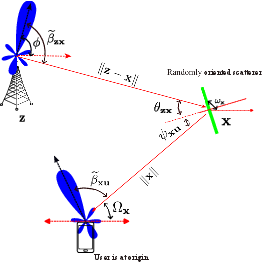}\\
{\bf \footnotesize (a)}
\end{minipage}
\begin{minipage}{0.21\textwidth}
	{\bf \footnotesize (b)}	\includegraphics[width=.9\linewidth,trim=0 .2 0 0,clip]{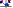}\\
			\includegraphics[width=.9\linewidth]{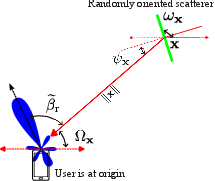}\vspace{-.1in}
			{\bf \footnotesize (c)}
\end{minipage}
			
\caption{Illustration showing (a) SE interference from a randomly oriented SE towards the typical UE located at origin, (b) BSs and various parameters with respect to the typical SE located at origin, (c) the scenario $\Omega_{\x}<\pi/2$ and $\Omega_{\x}<\omega_{\x}$.}
\label{fig:scattererscenrio1}
			\label{fig:basestationorientation1}
			\label{fig:scattererscenrio2}
			\vspace{-.1in}
		\end{figure}

We first  characterize the interference at the typical UE  from a single BS  due to a single  SE. As shown in Fig. \ref{fig:scattererscenrio1}(a), let us consider a BS located at $\z$, a SE located {at} $\x$ with random orientation and the UE located at the origin.
Let $\widetilde{\beta}_{\z\x}$ {denote} the angle between the axis of the maximum gain of the BS antenna {and} the direction of the SE. The angle $\theta_{\z\x}$ is between the normal to the SE to the direction of the BS and $\psi_{\x}$ is the angle between the normal to the SE orientation to the UE direction. Further, $\widetilde{\beta}_{\x}$ is the angle between the UE antenna's axis of maximum gain to the SE.
From Fig. \ref{fig:scattererscenrio1}(a), the power incident on the SE from the BS on the UE side  is
\begin{align*}
&P_{\z\x}=P_{\trm}g_{\trm}(\widetilde{\beta}_{\z\x})
\pathlossBStoSEs{\lstate{\z\x}}
(\|\z-\x\|)\cos (\theta_{\z\x})\mathrm{U}_{\z\x\mathbf{u}}.
\end{align*}Here, $\kappa_{\z\x}$ denotes the link state of $\z\x$,  ${\ell_{\kappa,\srm}}$ denotes the path-loss of the BS--SE link when it is LOS (${\kappa=\rmL}$) or NLOS (${\kappa=\rmN}$), $\lstate{l}\in\{\rmL,\rmN\} $ denotes the blocking state of link $l$ and $\mathrm{U}_{\z\x\mathbf{u}}$ is indicator RV denoting that the BS is {at} the same side of the SE as the UE. The power $P_{\z\x}$ is scattered by the SE {with} a fraction of it fall{ing} on the UE. 
The power incident on the UE located at the origin is
\begin{align*}
&P_{\z\x\mathbf{u}}=S^{2}P_{\z\x}g_{\rrm}(\widetilde{\beta}_{\x})
\pathlossSEtoUs{\lstate{\x\ueb}}
(\|\x\|)
\sprof{\cos (\psi_{\x})}.
\end{align*} 
Now, if we consider all SEs and BSs, the total interference at the typical user is given as 
\begin{align}
I_{\srm}&=\sum\nolimits_{\z\in \Phi_{\brm}}\sum\nolimits_{\x\in \Phi_{\srm}}P_{\z\x\ueb}.\label{interference}
\end{align}
\noindent The expression for $I_\srm$ in \eqref{interference} can be split into two parts. Let $P_\x$ denote the total power received at the SE at $\x$ from all BSs  \ie $P_{\x}=\sum\nolimits_{\z\in \Phi_{\brm}}P_{\z\x}$,  the interference $I_{\srm}$ at the UE can be written as
\begin{align}
I_\srm&=\sum\nolimits_{\x\in \Phi_{\srm}}S^{2}P_{\x}g_{\rrm}(\widetilde{\beta}_{\x})
\pathlossSEtoUs{\lstate{\x\ueb}}
(\|\x\|)
\sprof{\cos (\psi_{\x})}
\label{eq:IsMain}
\end{align}
The above representation splits the sum interference into two {terms,} the received power at each SE and the fraction of power received at UE from each SE. The LT of  $I_{\srm}$ is $\laplace_{I_{\srm}}(s)=\expect{\expS{-s I_{\srm}}}=$
\begin{align}
\hspace{-.08in}\expect{\prod\nolimits_{\x\in \Phi_{\srm}}\!\!\expS{\!-s S^{2}P_{\x}g_{\rrm}(\widetilde{\beta}_{\x})
\pathlossSEtoUs{\lstate{\x\ueb}}\!
(\|\x\|)
\sprof{\cos\psi_{\x}}
}
}.\label{eq:LT_1}
\end{align}{Further note that RVs} $P_\x$ and $P_{\x'}$ for two SEs at {$\x$ and $\x'$} have {the same BSs in their terms,} and hence, are correlated. Due to this correlation in  $P_{\x}$, it is difficult to simplify the LT further {as an application of the PGFL of a PPP requires independence across the marks}. Therefore, we present two approaches to approximate $I_{\srm}$ denoted as $I_{\srm,\app1}$ and $I_{\srm,\app2}$,  respectively. It is important to note that despite the correlation, the mean of $I_\srm$ can be computed exactly {as shown} in Section \ref{Sec:meanIs}.

\subsection{Approach with IID  $P_\x$'s Assumption} 
In this approach,  we consider that $P_\x$'s are IID across SEs  {$\x \in \Phi_\srm$}.
From \eqref{eq:LT_1}, we have 
the LT of  $I_{\srm}$ as $\laplace_{I_{\srm}}(s)=$
\begin{align}
&=\expect{\prod_{\x\in \Phi_{\srm}}\expS{-s S^{2}P_{\x}g_{\rrm}(\widetilde{\beta}_{\x})\ell_{\lstate{\x\ueb},\ue}(\|\x\|)\sprof{\cos (\psi_{\x})}
}
}\nonumber\\
		&\stackrel{(a)}{{\approx}}\mathbb{E}_{\Phi_\srm,\{\omega_\x\}}\left[\prod\nolimits_{\x\in \Phi_{\srm}}
		\mathbb{E}_{P_{\x}}\left[\exp\left(-sS^{2}P_{\x}g_{\rrm}(\widetilde{\beta}_{\x})\right.\right.\right.\nonumber\\
		&\qquad\left.\left.\left.\vphantom{\frac{num}{den}}\ell_{\lstate{\x\ueb},\ue}(\|\x\|)
		\sprof{\cos (\psi_{\x})}
		\right)\right]
		\right]\nonumber\\
		&\stackrel{(b)}=\expects{{\Phi_\srm,\{\omega_\x\}}}
		{\prod\nolimits_{\x\in \Phi_{\srm}}\!\!\!\mathcal{L}_{P_{\x}}(sS^{2}g_{\rrm}(\widetilde{\beta}_{\x})\vphantom{\frac{num}{den}}\ell_{\lstate{\x\ueb},\ue}
			(\|\x\|)
			\sprof{\cos (\psi_{\x})}
			)}\nonumber\\
		&\stackrel{(c)}=
		\exp\left(-
		\lambda_{\srm}\int_{\x\in \R^{2}}
		\left(1-\mathbb{E}_{\omega_\x}\left[\laplace_{P_{\x}\vert \omega_\x}(sS^{2}g_{\rrm}(\widetilde{\beta}_{\x}) \right.\right.\right.\nonumber\\
		&\left.\left.\left.
		\quad
		\ell_{\lstate{\x\ueb},\ue}(\|\x\|)
		\sprof{\cos (\psi_{\x})}
		)\right]\right)
		\losprob{\|\x\|}{\lstate{\x\ueb}}
		\dv \x\right)\label{eq:LTISApp2-1}
	\end{align}
	where step $(a)$ is {due to the} IID assumption of $P_\x$'s,  $(b)$ is from the definition of LT of $P_{\x}$ and step $(c)$ is obtained using the PGFL of PPP $\Phi_{\srm}$. To proceed further, we require LT of $P_\x$ which we obtain next.

	\subsubsection{Distribution of $P_\x$} 
	To compute the LT of $P_\x$, let us consider the typical SE at $\x_{\ob}$ at the origin as shown in Fig. \ref{fig:basestationorientation1}(b). 
	Since the distribution of $P_\x$ is not affected by $\omega_\x$, without loss of generality, we assume that the scattering face of the SE is oriented towards the positive ${\sf y}$-axis. 
Let $\delta_{\z_i}$ denote the angle from positive $\mathsf{x}$-axis and the direction of the $i$th BS with respect to the SE.  Therefore, the SE  scatters the incident power from the BSs that are located in the positive-${\sf y}$ half plane that is $\mathrm{U}_{\z_i\x_\ob\ueb}=1 \Leftrightarrow \delta_{\z_i} \geq0$. Let us denote BSs of  $\Phi_{\brm}$, $\Phi_{\rmL,\brm}$ and $\Phi_{\rmN,\brm}$ lying in the positive-$\mathsf{y}$ half plane by $\Phi_{\brm}^+$, $\Phi_{\rmL,\brm}^+$ and $\Phi_{\rmN,\brm}^+$. Hence, 
the total incident power $P_{\x_\ob}$ on the SE $\x_\ob$ is
	\begin{align}\label{incidentpowerscatter}
		\textstyle P_{\x_\ob}=&\sum\nolimits_{\z_{i}\in \Phi_{\rmL,\brm}^+}P_{\trm}
		\pathlossBStoSEs{\rmL}
		(\|\z_{i}\|)
		\cos(\theta_{\z_i})g_{\trm}(\widetilde{\beta}_{\z_i})\nonumber\\
		&\textstyle+\sum\nolimits_{\z_{i}\in\Phi_{\rmN,\brm}^+}P_{\trm}\ell_{\rmN,\srm}(\|\z_{i}\|)\cos(\theta_{\z_i})g_{\trm}(\widetilde{\beta}_{\z_i}),
	\end{align}
where $\widetilde{\beta}_{\z_{i}}$ is the angle between the beam-orientation of the antenna of the $i$th BS $\z_i$ and the typical SE, and $\theta_{\z_i}$ is the angle between the normal to the SE and the direction of the $i$th BS with respect to the SE. Here, the two terms in \eqref{incidentpowerscatter} denote the power from BSs that are LOS and NLOS from the SE.   Note that $\theta_{\z_i}=\left({\delta_{\z_i}}-{\pi}/{2}\right) \Rightarrow  \cos(\theta_{\z_i})=\sin(\delta_{\z_i})$
Further, let $\phi_{\z_i}$ denote the angle of the antenna-beam  at $i$th BS $\z_i$ which can be seen as the mark for each BS in $\Phi_{\brm}$ with uniform distribution in $(0,2\pi)$. The beam angle $\widetilde{\beta}_{\z_{i}}$ of the $i$th BS is {(see Appendix~B of the supplementary material \cite{pandey2026supplementary} for proof)}
	\begin{align}
		\widetilde{\beta}_{\z_{i}}({\delta}_{\z_i},{\phi}_{\z_i})=\begin{cases}
			\pi-|\delta_{\z_i}-\phi_{\z_i}|,& 0<\phi_{\z_i}<\pi\\
			|\delta_{\z_i}-(\phi_{\z_i}-\pi)|, & \pi<\phi_{\z_i}<2\pi.
		\end{cases}\label{tildebetapibypi}
	\end{align}

	Now, the LT of $P_{\x_\ob}$ is presented in the following theorem. 
	\begin{theorem}\label{thm4}
		The LT of $P_{\x_\ob}$ is 
		\begin{align*}
			&\laplace_{P_{\x_\ob}}(s)=	
			\laplace_{P_{\x_\ob,\rmL}}(s)\cdot	
			\laplace_{P_{\x_\ob,\rmN}}(s),\\
			&\text{where}\quad
			\laplace_{P_{\x_\ob,\lstate{}}}(s)=
			\exp\left(-\lambda_{\brm}\int_{0}^{\infty}
			\int_{\delta=0}^{\pi}
			\sum\nolimits_{i=1}^{2}\pt_{\trm,i}
			\right.\\
			&\left.\vphantom{\frac{num}{den}}
			\left(1-{\expS{-sP_{\trm}\Grm_{\trm,i}
			\pathlossBStoSEs{\lstate{}}
			(z)
			\sin (\delta)}}\right)
			\losprob{z}{\lstate{}}
			z \dv z \dv \delta\right)
		\end{align*}
		where
		$\Grm_{\trm,1}=\Grm_{\mrm,\trm}$ and $\Grm_{\trm,2}=\Grm_{\srm,\trm}$ denote the main and side lobe gain of the BS transmit antenna  with $\pt_{\trm,1}=\pt_{\trm}$ and $\pt_{\trm,2}=1-\pt_{\trm}$ as corresponding probabilities.
	\end{theorem}

\begin{proof}

From \eqref{incidentpowerscatter}, the LT $\laplace_{P_{\x_\ob}}(s)$ of $P_{\x_\ob}$ is
			\begin{align*}
				&=\expect{\expS{-s\sum\nolimits_{\z_{i}\in \Phi_{\brm}^+}P_{\trm}g_{\trm}(\widetilde{\beta}_{\z_i})\ell_{\lstate{\z_i},\srm}(\|\z_{i}\|)\cos (\theta_{\z_i})}}\\
				&=\prod_{\lstate{}\in\{\rmL,\rmN\}}\expect{
				\expS{-s\!\!
				\sum_{\z_{i}\in \Phi_{\lstate{},\brm}^+}
				\!\!\!
				P_{\trm}g_{\trm}(\widetilde{\beta}_{\z_i})
				\ell_{\lstate{},\srm}(\|\z_{i}\|)
				\cos (\theta_{\z_i})}}
			\end{align*}
		where the last step is due to independence of LOS and NLOS BSs. Let us focus on the LT of the first factor $\lstate{}=\rmL$ denoting the LOS BSs given as $\laplace_{P_{\x_\ob,\rmL}}(s)=$
			\begin{align*}
			&
				\expect{\expS{-\sum\nolimits_{\z_{i}\in \Phi_{\rmL,\brm}^+}sP_{\trm}g_{\trm}(\widetilde{\beta}_{\z_i})
				\pathlossBStoSEs{\rmL}
				(\|\z_{i}\|)
				\cos (\theta_{\z_i})}}\\
				&=\exp\left(-\lambda_{\brm}\int_{0}^{\infty}\int_{\delta_\z=0}^{\pi}\left(1-\right.\right.\\
				&\qquad \left.\left.\expects{\phi_{\z}}{e^{-sP_{\trm}g_{\trm}(\widetilde{\beta}_\z)
				\ell_{\rmL,\srm}
				(z)
				\sin (\delta_\z)}}\right)e^{-\beta z}z \dv z \dv \delta_\z\right),
			\end{align*}
			where the last step is due to PGFL of PPP $\Phi_{\rmL,\brm}^+$ and substituting the value of $\theta_{\z_i}$ from previous page. 
Now, owing to periodicity (wrapping) of angles over $2\pi$ and from \eqref{tildebetapibypi}, $g_\trm(\widetilde{\beta}_{\z})$ has the same distribution as $g_\trm(\phi_{\z})$ which 
takes $\Grm_{\mrm,\trm}$ and $\Grm_{\srm,\trm}$ with probability $\pt_{\trm}$ and $1-\pt_{\trm}$.
The proof for NLOS component is similar. 
\end{proof}

{ \subsubsection{Distribution of SE Interference}
Now, under IID $P_\x$ assumption, the LT of $ P_{\x}$ falling on each SE is  taken equal to the LT of $P_{\x_\ob}$ given by $\laplace_{P_{\x_\ob}}$.
		Now, $\psi_{\x}$ is the scattering angle from the SE at $\x$
		to the typical UE given as
			\begin{align}
				\cos\psi_{\x}=|\sin(\Omega_{\x}-\omega_{\x})|,\quad 0<\Omega_{\x}<2\pi,\label{psiequation}
		\end{align}
with  the orientation of the SE at $\x$ be $\omega_{\x}$ and its angular location be $\Omega_\x$ (see proof in Appendix \ref{app:psi_xproof}). Further, let $\widetilde{\beta}_{\x}$ be the angle between the beam-orientation of the UE antenna and the SE $\x$ direction given as $\widetilde{\beta}_{\x}=\phi_{\ueb}-\Omega_\x$ where  $\phi_{\ueb}$ denote the angle of the UE antenna-beam with uniform distribution in $(0,2\pi)$.  Without loss of generality, we can take $\phi_\ueb=0$.
 Now, we present the LT of $I_\srm$   in the following Theorem. 
\begin{theorem} \label{LTinterferecenapp2}
			Under IID $P_\x$ assumption, the LT of $I_{\srm}$ can be approximated as
			\begin{align*}
				&\laplace_{I_{{\srm,\app2}}}(s)=	\laplace_{I_{\srm,\rmL,\app2}}(s)\cdot\laplace_{I_{\srm,\rmN,\app2}}(s),\\
				&\text{where }\textstyle\laplace_{I_{\srm,\lstate{},\app2}}(s)=\exp\left(-\sum\nolimits_{i=1}^{2}
				{2\pi \pt_{\rrm,i}\lambda_{\srm}}\int_{x=0}^{\infty}\int_{y=0}^{1}\right.\\
				&\textstyle\left.\vphantom{\frac{num}{den}}\left(1-\laplace_{P_{\x_\ob}}\left(sS^{2}{\Grm_{\rrm,i}}
				\pathlossSEtoUs{\lstate{}}
				(x)
				\sprof{y}\right)\right)\PDFY(y)x
				\losprob{x}{\lstate{}}
				\dv y   \dv x\right),
			\end{align*}
where $\laplace_{P_{\x_\ob}}(s)$ and $v(y)$ are given in Theorem~\ref{thm4} and Proposition~\ref{propos2}, respectively. Gain  $\Grm_{\rrm,1}=\Grm_{\mrm,\rrm}$, $\Grm_{\rrm,2}=\Grm_{\srm,\rrm}$ denote main and side gain of receive antenna with  $\pt_{\rrm,1}=\pt_{\rrm}$ and $\pt_{\rrm,2}=1-\pt_{\rrm}$ as the corresponding probabilities. 
\end{theorem}}

{\begin{proof}
Noticing that the LT of $P_\x$ is independent of $\omega_\x$, from \eqref{eq:LTISApp2-1}, we get 
\begin{align}
&\textstyle\laplace_{I_{{\srm,\app2}}}(s)
		=
		\exp\left(-
		\lambda_{\srm}\int_{\x\in \R^{2}}
		\left(1-\mathbb{E}_{\omega_\x}\left[\right.\right.\right.\nonumber\\
		\!\!\!&\textstyle\left.\left.\left.
		\laplace_{P_{\x_\ob}}(sS^{2}g_{\rrm}(\widetilde{\beta}_{\x})
		\pathlossSEtoUs{\lstate{\x\ueb}}
		(\|\x\|)
		\sprof{\cos (\psi_{\x})}
		)\right]\right)
		\losprob{\|\x\|}{\lstate{\x\ueb}}
		\dv \x\right).\nonumber%
\end{align}
From \eqref{psiequation} and Proposition \ref{propos2}, we have
\begin{align}
&\textstyle\laplace_{I_{{\srm,\app2}}}(s)
		=
		\exp\left(-
		\lambda_{\srm}\int_{\x\in \R^{2}}
		\int_{y=0}^1
		\left(1-\right.\right.\nonumber\\
		&\textstyle\left.\left.
		\laplace_{P_{\x_\ob}}(sS^{2}g_{\rrm}(\widetilde{\beta}_{\x})
		\ell_{\lstate{\x\ueb},\ue}(\|\x\|)
		\sprof{y}
		)
		\right)\PDFY(y)
		\losprob{\|\x\|}{\lstate{\x\ueb}}\dv y
		\dv \x\right).\nonumber
\end{align}
Now, noticing that $\widetilde{\beta}_{\x}=\phi_{\ueb}-\Omega_\x=-\Omega_\x$,  we get
\begin{align}
	&\laplace_{I_{{\srm,\app2}}}(s)
	=
	\exp\left(-\sum\nolimits_{i=1}^2 2\pi \pt_{\rrm,i}
	\lambda_{\srm}\int_{x=0}^{\infty}
	\int_{y=0}^1
	\left(1-\right.\right.\nonumber\\
	&\left.\left.\!\!\!\!\!
	\laplace_{P_{\x_\ob}}(sS^{2}\Grm_{\rrm,i}
	\ell_{\lstate{\x\ueb},\ue}(x)
	\sprof{y}
	)
	\right)\PDFY(y)x
	\losprob{x}{\lstate{\x\ueb}}\dv y
	\dv x\right)\label{eq:LTISApp2-4}.
\end{align}
Now spli{t}ting it over LOS and NLOS SEs gives the desired result.
\end{proof}Now, we will provide the LT of $I_\srm$ at $s=-ju$ to be used later for evaluation of CP as given in the following Corollary. 
\begin{corollary}\label{Cor:LTIsIID}
The LT  of $I_{{\srm,\app2}}$ at $s=-ju$ is given as
\begin{align}
&\laplace_{I_{\srm,\app2}}(-ju)=R_{\srm,\app2}(u)\expS{j\Theta_{\srm,\app2}(u)}\quad\text{with}\nonumber\\
&R_{\srm,\app2}(u)= R_{\srm,\rmL,\app2}(u) R_{\srm,\rmN,\app2}(u),
\Theta_{\srm,\app2}(u)=\Theta_{\srm,\rmL,\app2}(u)+\!\Theta_{\srm,\rmN,\app2}(u),\nonumber
\\	
		&\text{where } R_{\srm,\lstate{},\app2}(u)=\nonumber
		\\
		&\!\exp\left(-\sum_{i=1}^{2}{\pt_{\rrm,i}2\pi\lambda_{\srm}}\int_{0}^{\infty}\!\!\!\int_{y=0}^{1}\!\!\!
		\left(1-R_{\x_\ob}(
		uS^{2}{\Grm_{\rrm,i}}
		\pathlossSEtoUs{\lstate{}}(x)
		\sprof{y}
		)
		\right.\right.\nonumber\\
		&\ \  \left.\left.\cos\left(\Theta_{\x_\ob}
		(
		uS^{2}{\Grm_{\rrm,i}}
		\pathlossSEtoUs{\lstate{}}(x)
		\sprof{y}
		)\right) \PDFY(y)x
		\losprob{x}{\lstate{}}
		\dv y
		   \dv x\right)\right)\label{RIsLapp2}\\
		&\Theta_{\srm,\lstate{},\app2}(u)=\nonumber\\
		&\sum\nolimits_{i=1}^{2}
		{\pt_{\rrm,i}2\pi\lambda_{\srm}}
		\int_{x=0}^{\infty}\int_{y=0}^{1}R_{\x_\ob}(
		uS^{2}{\Grm_{\rrm,i}}
		\pathlossSEtoUs{\lstate{}}(x)
		\sprof{y}
		)
		\nonumber\\
		&
		\quad \sin\left(\Theta_{\x_\ob}(
		uS^{2}{\Grm_{\rrm,i}}
		\pathlossSEtoUs{\lstate{}}(x)
		\sprof{y}
		)\right)\PDFY(y)x
		\losprob{x}{\lstate{}}
		\dv y   \dv x.\label{TIsLapp2}
		\end{align}
Here, $R_{\x_\ob}(u)$ and $\Theta_{\x_\ob}(u)$ denote the magnitude and argument of the LT of $P_{\x_\ob}$ at $s=-ju$ and are given as (for proof, see Appendix~\ref{app:corIID})
		\begin{align}
 &R_{\x_\ob}(u)=R_{\x_\ob,\rmL}(u)R_{\x_\ob,\rmN}(u),\nonumber\\
 &\, \Theta_{\x_\ob}(u)=\Theta_{\x_\ob,\rmL}(u)+\Theta_{\x_\ob,\rmN}(u),\nonumber\\
			&R_{\x_\ob,\lstate{}}(u)=\label{rsl}\\
			&\expU{
			-\lambda_{\brm}\int_{0}^{\infty}\int_{\delta=0}^{\pi}\sum\nolimits_{i=1}^{2}\pt_{\trm,i}\left(1-
			{\left(\cos\left(uP_{\trm}\Grm_{\trm,i}
			\pathlossBStoSEs{\lstate{}}(z)
			\sin (\delta)\right)\right)}\right)
			\losprob{z}{\lstate{}}
			z \dv z \dv \delta
			},\nonumber\\
			&\textstyle\Theta_{\x_\ob,\lstate{}}(u)=\lambda_{\brm}\int_{0}^{\infty}\int_{\delta=0}^{\pi}\sum\nolimits_{i=1}^{2}\pt_{\trm,i}\sin\left(uP_{\trm}\Grm_{\trm,i}
			\pathlossBStoSEs{\lstate{}}(z)
			\right.\nonumber\\
			&\qquad\left.\sin (\delta)\right)
			\losprob{z}{\lstate{}}
			z \dv z \dv \delta\label{thetasl}.
		\end{align}
\end{corollary}}

\subsection{Approach with Mean $P_{\x}$ Assumption} 
We now present another {but simpler approach} where we approximate $P_{\x}$ by its mean $m_{\ob}$ denoting the mean power incident on the typical SE from all BS{s}.
\subsubsection{Mean incident power at the SE}
To compute $m_{\ob}$, we can consider the typical SE at $\x_{\ob}$ at the origin as shown in Fig. \ref{fig:basestationorientation1}(b) without loss of generality.
Now, we present the mean value {of} $P_{\x_\ob}$ in the following Theorem.

\begin{theorem}\label{thm:ms}
			The average of incident power $P_{\x_\ob}$ from all the BSs to the typical SE at $\x_\ob$ is (for proof, see Appendix \ref{app:thm:ms}) 
			\begin{align}
			&m_{\ob}=\expect{P_{\x_\ob}}=\expect{P_{\x_\ob,\rmL}}+\expect{P_{\x_\ob,\rmN}}=\nonumber\\
				&\!\!2P_{\trm}
				\widetilde{g_\trm}\lambda_{\brm}%
				\left(\int_{0}^{\infty}
				\!\!\!\left(e^{-\beta z}
				\pathlossBStoSEs{\rmL}(z)
				+\!
				(1-e^{-\beta z})\ell_{\rmN,\srm}(z)\right)z\dv z\!\right),\label{Psapp1}
			\end{align}
			where $\widetilde{g_\trm}=\left(\Grm_{\mrm,\trm}\pt_{\trm}+\Grm_{\srm,\trm}(1-\pt_\trm) \right)$ denotes the average radiated power from the transmit antenna. 
\end{theorem}

\subsubsection{Distribution of $I_\srm$}
Now, approximating $ P_{\x}$ falling on each SE to be fixed and equal to the mean $m_{\ob}$,   from \eqref{eq:IsMain}, the interference at the typical UE {is}
	\begin{align}\label{Is1}
	\!\!\!\!	I_{\srm,\app1}=\sum\nolimits_{\x_{i}\in \Phi_{\srm}}S^{2}m_{\ob}g_{\rrm}(\widetilde{\beta}_{\x_{i}})
		\pathlossSEtoUs{\lstate{\x_i\ueb}}
		(\|\x_i\|)\sprof{\cos(\psi_{\x_i})}.
	\end{align}
Now we present the LT of $I_\srm$ in the following Theorem.
\begin{theorem}\label{theorem 4}
	The LT of the interference at the typical user due to all SEs {under the mean $P_{\x}$ approximation} is given as (for proof, see Appendix \ref{app:D})
	\begin{align}
		&\laplace_{I_{\srm,\app1}}(s)=	\laplace_{I_{\srm,\rmL,\app1}}(s)\times\laplace_{I_{\srm,\rmN,\app1}}(s),\label{LaplaceIsapp1}\\
		&\text{where}\quad\laplace_{I_{\srm,\rmL,\app1}}(s)=\exp\left(-\sum\nolimits_{i=1}^{2}2\pi\pt_{\rrm,i}{\lambda_{\srm}}\int_{x=0}^{\infty}\int_{y=0}^{1}\right.\nonumber\\
		&\left.\left(1-e^{-s S^{2}m_{\ob}\Grm_{\rrm,i}
			\pathlossSEtoUs{\rmL}(x)
			\sprof{y}}\right)e^{-\beta x} x \PDFY(y)\dv y\dv x\right),\label{LaplaceIsapp1L}\\
		&\laplace_{I_{\srm,\rmN,\app1}}(s)=\exp\left(-\sum\nolimits_{i=1}^{2}2\pi\pt_{\rrm,i}{\lambda_{\srm}}\int_{x=0}^{\infty} \int_{y=0}^{1}\left(1-\vphantom{\frac{num}{den}}\right.\right.\nonumber\\
		&\left.\left.e^{-s S^{2}m_{\ob}\Grm_{\rrm,i}
			\ell_{\rmN,\ue}(x)
			\sprof{y}}
		\right) x(1-e^{-\beta x}) \PDFY(y)\dv y \dv x\right),\label{LaplaceIsapp1N}
	\end{align}
	where $\Grm_{\rrm,1}=\Grm_{\mrm,\rrm}$, $\Grm_{\rrm,2}=\Grm_{\srm,\rrm}$ denote main and side gain of receive antenna with  $\pt_{\rrm,1}=\pt_{\rrm}$ and $\pt_{\rrm,2}=1-\pt_{\rrm}$ as the corresponding probabilities.
\end{theorem}We also compute the LT of $I_{\srm,\app1}$ at $s=-ju$
as provided in the following Corollary to derive the coverage later.
{\begin{corollary}\label{cor5.1}
	The LT of interference $I_{\srm,\app1}$ at $s=-ju$ is
		\begin{align}
		&\laplace_{I_{\srm,\app1}}(-ju)=R_{\srm,\app1}(u) \expU{j\Theta_{\srm,\app1}(u)}\\
			&\text{with }R_{\srm,\app1}(u)= R_{\srm,\rmL,\app1}(u) 
						R_{\srm,\rmN,\app1}(u),\,\\
			&\Theta_{\srm,\app1}(u)=\Theta_{\srm,\rmL,\app1}(u)
			+\Theta_{\srm,\rmN,\app1}(u)\\
&R_{\srm,\lstate{},\app1}(u)=\exp\left(-\sum\nolimits_{i=1}^{2}2\pi\pt_{\rrm,i}{\lambda_{\srm}}\int_{x=0}^{\infty}\int_{y=0}^{1}\left(1-\right.\right.\nonumber\\
		&\quad\left.\left.\cos\left(u S^{2}m_{\ob}\Grm_{\rrm,i}
		\pathlossSEtoUs{\lstate{}}(x)
		\sprof{y}
		\right)\right) 
		\losprob{x}{\lstate{}}
		x \PDFY(y)\dv y\dv x\right)\label{RIsLapp1}\\
		&\Theta_{\srm,\lstate{},\app1}(u)=\sum_{i=1}^{2}2\pi\pt_{\rrm,i}{\lambda_{\srm}}\int_{0}^{\infty}\!\!\int_{0}^{1}\nonumber\\&\qquad\sin\left(u S^{2}m_{\ob}\Grm_{\rrm,i}
		\pathlossSEtoUs{\lstate{}}(x)
		\sprof{y}
		\right)
		\losprob{x}{\lstate{}}
		x \PDFY(y)\dv y\dv x\label{TIsLapp1}.
\end{align}
	\end{corollary}
\begin{proof}
	For LOS component, substituting $s\!=\!-ju$ in  \eqref{LaplaceIsapp1L} gives 
	\begin{align}
		&\laplace_{I_{\srm,\rmL,\app1}}(-ju)=\exp\left(-\sum\nolimits_{i=1}^{2}2\pi\pt_{\rrm,i}{\lambda_{\srm}}\int_{x=0}^{\infty}\int_{y=0}^{1}\right.\nonumber\\
		&\left.\left(1-e^{ju S^{2}m_{\ob}\Grm_{\rrm,i}\ell_{\rmL,\ue}(x)\sprof{y}}\right)e^{-\beta x} x \PDFY(y)\dv y\dv x\right)\nonumber\\
		&=R_{\srm,\rmL,\app1}(u)\expS{j \Theta_{\srm,\rmL,\app1}(u)},\nonumber	
	\end{align}
	where $R_{\srm,\rmL,\app1}(u)$ ($\Theta_{\srm,\rmL,\app1}(u)$) are the magnitude (argument) of   $\laplace_{I_{\srm,\rmL,\app1}}(-ju)$.
	The proof for NLOS component is similar.
\end{proof}}

\subsection{Mean SE Interference}\label{Sec:meanIs}
We now present the mean scattered interference from all BSs. From \eqref{eq:IsMain}, the mean interference $I_{\srm}$ at the UE can be written as
\begin{align}
&\expect{I_\srm}=\expect{\sum\nolimits_{\x\in \Phi_{\srm}}S^{2}P_{\x}g_{\rrm}(\widetilde{\beta}_{\x})
\pathlossSEtoUs{\lstate{\x\ueb}}
(\|\x\|)
\sprof{\cos (\psi_{\x})}}\nonumber\\
&=\expect{\sum\nolimits_{\x\in \Phi_{\srm}}\!\!\!S^{2}m_\ob g_{\rrm}(\widetilde{\beta}_{\x})
\pathlossSEtoUs{\lstate{\x\ueb}}
(\|\x\|)
\sprof{\cos (\psi_{\x})}}.\label{eq:meanIs1}
\end{align}

\begin{theorem}\label{theorem2}
		The mean SE interference at the typical UE  is 
		\begin{align*}
			&\expect{I_{\srm}}={2S^{2}m_{\ob}\lambda_{\srm}\widetilde{g_\rrm}}\\
			&\times \left(\int_{x=0}^{\infty} (e^{-\beta x}\ell_{\rmL,\ue}(x)%
			+
			(1-e^{-\beta x})\ell_{\rmN,\ue}(x))
			x\dv x\right),
		\end{align*}
		with $\widetilde{g_\rrm}=\frac{\Grm_{\mrm,\rrm}\gamma_{\rrm}+\Grm_{\srm,\rrm}(2\pi-\gamma_{\rrm})}{2\pi}$ 
		where $m_\ob$ is given in \eqref{Psapp1}. 
\end{theorem}	
\begin{proof}
		From \eqref{eq:meanIs1}, we have
		\begin{align*}
		&\expect{I_{\srm}}\stackrel{(a)}=\expect{\sum\nolimits_{\x_{i}\in \Phi_{\rmL,\srm}}S^{2}m_{\ob}g_{\rrm}(\widetilde{\beta}_{\x_{i}})\ell_{\rmL,\ue}(\|\x_i\|)
		\sprof{\cos(\psi_{\x_i})}\right.\\
			&\left.+\sum\nolimits_{\x_{i}\in \Phi_{\rmN,\srm}}S^{2}m_{\ob}g_{\rrm}(\widetilde{\beta}_{\x_{i}})\ell_{\rmN,\ue}(\|\x_i\|)\sprof{\cos(\psi_{\x_i})}},
	\end{align*}
where  step $(a)$ is obtained by breaking the summation into LOS and NLOS links. Let us consider the LOS component first.
\begin{align*}
&\expect{I_{\srm,\rmL}}=\expect{\sum\nolimits_{\x_{i}\in \Phi_{\rmL,\srm}}S^{2}m_{\ob}g_{\rrm}(\widetilde{\beta}_{\x_{i}})\ell_{\rmL,\ue}(\|\x_i\|)
\sprof{\cos(\psi_{\x_i})}}\\
&\stackrel{(a)}=\expects{ \Phi_{\rmL,\srm}}
	{\sum\nolimits_{\x_{i}\in \Phi_{\rmL,\srm}}S^{2}
	m_{\ob}
	\expects{\phi_{\ueb}}
	{g_{\rrm}(\widetilde{\beta}_{\x_{i}})}
	\ell_{\rmL,\ue}(\|\x_i\|)
	\sprof{\cos \psi_{\x_i}}}\\
&\stackrel{(b)}=\expects{ \Phi_{\rmL,\srm}}
	{\sum\nolimits_{\x_{i}\in \Phi_{\rmL,\srm}}S^{2}
	m_{\ob}
	\widetilde
	{g_{\rrm}}
	\ell_{\rmL,\ue}(\|\x_i\|)\expects{\omega_{\x_i}}{
	\sprof{\cos \psi_{\x_i}}}},
	\end{align*}
where step $(a)$ is obtained by law of total probability, and step (b) is due to taking the expectation with respect to uniformly distributed $\phi_{\ueb}$.
Now, from \eqref{psiequation} and \eqref{eq:prop1}, $\expects{\omega_{\x_i}}{\sprof{\cos \psi_{\x_i}}}=1/\pi$.  Substituting it back, we get
\begin{align*}	
&\expect{I_{\srm,\rmL}}=\expects{ \Phi_{\rmL,\srm}}
	{\sum\nolimits_{\x_{i}\in \Phi_{\rmL,\srm}}S^{2}
	m_{\ob}
	\widetilde
	{g_{\rrm}}
	\ell_{\rmL,\ue}(\|\x_i\|)\frac1\pi}
	\\
	&\stackrel{(c)}=
	\int_{x=0}^{\infty}\int_{\Omega=0}^{2\pi}S^{2}m_{\ob}\lambda_{\srm}\widetilde
	{g_{\rrm}}
	\ell_{\rmL,\ue}(\|\x\|)\frac1\pi e^{-\beta x}
x \dv \Omega \dv x\\
&=
	\frac1\pi 2\pi \int_{x=0}^{\infty}S^{2}m_{\ob}\lambda_{\srm}\widetilde
	{g_{\rrm}}
	\ell_{\rmL,\ue}(\|\x\|) e^{-\beta x}
x  \dv x,
\end{align*}where  step $(c)$ is due to Campbell's theorem \cite{andrews2023introduction}.
Applying similar steps for NLOS and summing the two, we get the desired result.
	\end{proof}

\section{Characterization of Interference due to BSs}
We now characterize the interference due to all BSs at the typical UE. This includes the interference caused by LOS and NLOS BSs and received at both main and side lobes of the UE receiver antenna. {Before proceeding,} we first present the distribution of the serving distance $R_\ob$ in the following.
\begin{lemma}\label{lem:Ro}
{Conditioned on the typical UE being connected to a LOS BS, the} PDF of the serving distance is given as {$g_{R_\ob}(r)=f_{R_\ob}(r)/A_{\rmL}$ with}
\begin{align}
f_{R_\ob}(r)&=2\pi{\lambda_\brm r}\losprob{r}{\rmL}e^{-2\pi  \lambda_\brm \int_0^r\losprob{u}{\rmL} u\dd u }\nonumber\\
&= 2\pi \lambda_{\brm}r e^{-\beta r}e^{-\frac{2\pi  \lambda_\brm}{\beta^2} \left(1 - e^{-\beta r}(1 + \beta r)\right)}\label{eq:RoPDF}
\end{align}
where, $A_{\rmL}$ is the probability that the typical UE has at least one LOS BS given as
\begin{align} \hspace{-.2in}
A_{\rmL}\!=\!\int_{0}^{\infty}\!\!\!\!f_{R_\ob}(r)\dd r\!=\!
1-\expU{-2\pi\lambda_\brm\!\int_0^\infty\! \losprob{u}{\rmL}u\dd u}
\!=\!1\!-\!e^{-2\pi\frac{\lambda_\brm}{\beta^{2}}}.\!\!\!\label{eq:AL}
\end{align}
\end{lemma} 
Note that with probability $1-A_{\rmL}$, there may not exist any LOS BS from the typical UE, in which case it is not served. The coverage  should be averaged with respect to $f_{R_\ob}(r)$, so that a UE without a LOS BS is counted as being in outage. Let $I_{\brm,\rmL}$ and $I_{\brm,\rmN}$ denote the interference due to LOS and NLOS BSs.
Therefore, the BS interference at the typical UE is
\[	I_{\brm}=I_{\brm,\rmL}+I_{\brm,\rmN}
\text{ with } I_{\brm,\lstate{}}=\sum\nolimits_{\z\in \Phi_{\lstate{},\brm}}
g_{\z}P_{\trm}
\pathlossBStoUs{\lstate{}}
(\|\z\|).\]
Here, $g_{\z}$ denotes the combined transmit and receive antenna gain for the signal from a BS at $\z$ to the typical UE which takes values $\Grm_i$ with probabilities $\pt_i$ as given in \eqref{eq:AntennaGainValues}. 
In the following  Theorem, we present the
distribution of $I_{\brm}$.

\begin{theorem}\label{LT_of_Ib}
	The LT of $I_{\brm}$ conditioned that serving BS is located at a distance $R_{\ob}$ is (for proof, see Appendix \ref{proof_of_LT_of_Ib})
		\begin{align*}
		&\laplace_{I_{\brm}\vert R_{\ob}}(s)=
		\laplace_{I_{\brm,\rmL}\vert R_{\ob}}(s)	\laplace_{I_{\brm,\rmN}\vert R_{\ob}}(s),\\
		&\text{where }\textstyle\laplace_{I_{\brm,\rmL}\vert R_{\ob}}(s)
		=\exp\left(-2\pi \sum\nolimits_{i=1}^{4} 
		\int_{R_{\ob}}^{\infty}\lambda_{\rmL,\brm}(z)
		\pt_{i}\right.\nonumber\\
		&\vphantom{\frac{num}{den}}\left.\left(1-\expS{-sP_{\trm}\Grm_{i}
		\pathlossBStoUs{\rmL}
		(z)}\right) z\dv z\right),
		\\
	&\textstyle\laplace_{I_{\brm,\rmN}\vert R_{\ob}}(s)	=\exp\left(-2\pi \sum\nolimits_{i=1}^{4} \int_{0}^{\infty}\lambda_{\rmN,\brm}(z)\pt_{i}\right.\\
	&\left.\left(1-\expS{-sP_{\trm}\Grm_{i}
	\pathlossBStoUs{\rmN}
	(z)}\right) z\dv z\right).
	\end{align*}
\end{theorem}

\noindent Here, it is important to note the difference of exclusion radi{i} for LOS and NLOS BSs \cite{andrews2023introduction}. Due to the cell associat{ion} policy, the LOS interfering BSs must lie outside the distance of $R_\ob$ while NLOS BSs can lie anywhere.
\begin{corollary}\label{average_Ib}
	The mean interference $I_{\brm}$ from all BSs at the typical UE is given as
	\begin{align*}
		\textstyle\mathbb{E}[I_{\brm}]=&
		\textstyle2\pi \lambda_{\brm} P_{\trm}\sum\nolimits_{i=1}^{4}\pt_{i}\Grm_{i} \left(\expects{R_\ob}{\int_{R_\ob}^{\infty}
		\losprob{z}{\rmL}
		\pathlossBStoUs{\rmL}{(z)}z\dv z}\right. \\
		&\textstyle\left.
		\qquad+\int_{0}^{\infty}\losprob{z}{\rmN}
		\pathlossBStoUs{\rmN}{(z)}
		z\dv z\right).
	\end{align*}
\end{corollary}
\begin{corollary}\label{cor6.1}
	The LT of interference $I_{\brm}$ at $s=-ju$ is
		\begin{align}
		&\laplace_{I_{\brm}\vert R_\ob}(-ju)=R_{\brm}(R_\ob,u)\expU{j\Theta_{\brm}(R_\ob,u)}\quad \text{where}\\
&R_{\brm}(r,u)=R_{\brm,\rmL}(r,u)\times R_{\brm,\rmN}(r,u),\label{Rbab}\\  
&\Theta_{\brm}(r,u)=\Theta_{\brm,\rmL}(r,u)+\Theta_{\brm,\rmN}(r,u),
\label{thetRbab}\\
&R_{\brm,\rmL}(r,u)=\expU{-2\pi \sum\nolimits_{i=1}^{4} \int_{r}^{\infty}\lambda_{\rmL,\brm}(z)\pt_{i}
\left(1-\cos\left(uP_{\trm}\Grm_{i}\pathlossBStoUs{\rmL}(z)\right)\right) z\dv z}\nonumber\\
&\Theta_{\brm,\rmL}(r,u)=\textstyle2\pi \sum\nolimits_{i=1}^{4} \int_{r}^{\infty} \lambda_{\rmL,\brm}(z)\pt_{i}\sin\left(uP_{\trm}\Grm_{i}\pathlossBStoUs{\rmL}(z)\right) z\dv z\nonumber\\
&R_{\brm,\rmN}(r,u)=\expU{-2\pi \sum\nolimits_{i=1}^{4} \int_{0}^{\infty}
\vphantom{\frac{num}{den}}\lambda_{\rmN,\brm}(z)
\pt_{i}\left(1-\cos\left(uP_{\trm}\Grm_{i}\pathlossBStoUs{\rmN}(z)\right)\right) z\dv z}\nonumber\\
&\Theta_{\brm,\rmN}(r,u)=\textstyle 2\pi \sum\nolimits_{i=1}^{4} \int_{0}^{\infty}\lambda_{\rmN,\brm}(z)\pt_{i}\sin\left(uP_{\trm}\Grm_{i}\pathlossBStoUs{\rmN}(z)\right) z\dv z.\nonumber
\end{align}
\end{corollary}

\section{Coverage Analysis}
Equipped with the LT of interferences from BSs and SEs, we now provide the  CP of the typical UE in the following Theorem (for proof, see Appendix \ref{proof_of_coverage_probablity_general}).  	
\begin{theorem}\label{coverage_probablity_general}
	The CP ${{\rm p_{c}}}(\tau)$ for the typical UE  is 
		\begin{align}
		{{\rm p_{c}}}(\tau)=&\frac{1}{2}-\frac{1}{\pi}\int_{r=0}^{\infty}\int_{u=0}^{\infty}\mathrm{Im}\left[\vphantom{\frac{e^{\frac{num}{den}}}{den}}e^{-ju {\mathcal{S}}(r)+ju\tau N_{\ob}}\right.\nonumber\\	
		&\left. \laplace_{I_\brm\vert R_\ob=r}(-ju{\tau})
		\laplace_{I_{{\srm}}}(-ju{\tau})
		\vphantom{\frac{num}{den}}\right]\frac{\dv u}{u}f_{R_{\ob}}(r)\dv r. \label{Lemm1eqn}
		\end{align}
		\end{theorem}\noindent{Utilizing the value of $\laplace_{I_{\srm,\app2}}(-ju{\tau})$ and $\laplace_{I_{\srm,\app1}}(-ju{\tau})$ from Corollary~\ref{Cor:LTIsIID} and Corollary~\ref{cor5.1} for the two approaches respectively in Theorem~\ref{coverage_probablity_general}, and then extracting the imaginary part followed by averaging with respect to $R_{\ob}$, we get the CP for the two approaches as given in the following two lemmas.}

\begin{lemma}
	The CP of the typical UE in a THz network {in presence of interference from BSs and scatterers} under {the} IID SE incident power {approximation} is given as
		\begin{align*}
			&{\rm p_{c,\app2}}(\tau)=\frac{1}{2}+\frac{1}{\pi}\int_{r=0}^{\infty}\int_{u=0}^{\infty}R_{\brm}(r,{\tau}u)R_{\srm,\app2}({\tau}u)\\
			&\sin\left(u {\mathcal{S}}(r)-\left(\Theta_{\srm,\app2}({\tau}u)+\Theta_{\brm}(r,{\tau}u)+u\tau N_{\ob}\right)\right)\frac{\dv u}{u}f_{R_{\ob}}(r)\dv r, 
		\end{align*}where $R_{\srm,\app2}(u)$, $\Theta_{\srm,\app2}(u)$, $R_{\brm}(r,u)$ and $\Theta_{\brm}(r,u)$ are provided in 
		Corr. \ref{Cor:LTIsIID}, \eqref{Rbab} and \eqref{thetRbab}, respectively.
\end{lemma}

\begin{lemma}
	The CP of the typical UE in a THz network {in presence of interference from BSs and scatterers} under {the} mean SE incident power approximation is given as
\begin{align*}
		&{{\rm p}_{{\rm c},\app1}}(\tau) =\frac{1}{2}+\frac{1}{\pi}\int_{u=0}^{\infty}\frac1u\nonumber\\
		&\qquad \mathbb{E}_{R_{\ob}}\left[H(R_{\ob},\tau u)\sin\left(u{\mathcal{S}}(R_{\ob})-\Theta(R_{\ob},\tau u)-u\tau N_{\ob}\right)\right] {\dv u},
		\intertext{where}
		&H(R_{\ob},\tau u)=e^{-{2}\pi \sum_{i=1}^{4} 
		\int_{R_{\ob}}^{\infty}\lambda_{\rmL,\brm}(z)
		\pt_{i}\left(1-\cos\left(u \tau P_{\trm}\Grm_{i}
		\pathlossBStoUs{\rmL}
		(z)
		\right)\right) z\dv z}\\
	&\qquad \times e^{-{2}\pi \sum_{i=1}^{4} 
		\int_{0}^{\infty}\lambda_{\rmN,\brm}(z)\pt_{i}
		\left(1-\cos\left(u \tau P_{\trm}\Grm_{i}
		\pathlossBStoUs{\rmN}(z)
		\right)\right) z\dv z}\\
	& \!\times\! e^{-\sum_{i=1}^{2}2\pt_{\rrm,i}
		{\lambda_{\srm}}\int_{0}^{\infty}\int_{0}^{\pi}
		\left(1-\cos\left(u {\tau} S^{2}m_{\ob}\Grm_{\rrm,i}
		\pathlossSEtoUs{\rmL}(x)
		{\sprof{\sin t}}\right)\right)
		xe^{-\beta x}{\dv t}\dv x}\\
	& \!\times\! e^{-\!\sum_{i=1}^{2}\!2\pt_{\rrm,i}{\lambda_{\srm}}\!
		\int_{0}^{\infty}\!\int_{0}^{\pi}
		\left(1-\!\cos(u {\tau} S^{2}m_{\ob}\Grm_{\rrm,i}
		\pathlossSEtoUs{\rmN}
		(x)
		{\sprof{\sin t}})\right) x(1-e^{-\beta x}){\dv t}\dv x}\\
	&\Theta(R_{\ob},\tau u)= \textstyle\left.{2}\pi \sum_{i=1}^{4} 
	\int_{R_{\ob}}^{\infty}\lambda_{\rmL,\brm}(z)\pt_{i}
	\left(\sin\left(u \tau P_{\trm}\Grm_{i}
	\pathlossBStoUs{\rmL}(z)
	\right)\right) 
	z\dv z\right.\\
	&\left. + {2}\pi \sum_{i=1}^{4} 
	\int_{{0}}^{\infty}
	\lambda_{\rmN,\brm}(z)\pt_{i}
	\sin\left(u \tau P_{\trm}\Grm_{i}
	\pathlossBStoUs{\rmN}(z)
	\right)z\dv z
	\!+\!\sum_{i=1}^{2}\pt_{\rrm,i}
	{2\lambda_{\srm}}\right. \\
	&\textstyle\left.\int_{x=0}^{\infty}
	\left(\int_{0}^{\pi}\sin\left(u {\tau} S^{2}m_{\ob}\Grm_{\rrm,i}
	\pathlossSEtoUs{\rmL}
	(x)
	{\sprof{\sin t}}\right){\dv t} \right) xe^{-\beta x}  \dv x\right.\\
	&\textstyle\left. +\sum_{i=1}^{2}\pt_{\rrm,i}{2\lambda_{\srm}}\int_{x=0}^{\infty}\!\left(\int_{0}^{\pi}\sin\left(u {\tau} S^{2}m_{\ob}\Grm_{\rrm,i}
	\pathlossBStoUs{\rmN}
	(x)
	{\sprof{\sin t}}\right){\dv t} \!\right)\right. \\
	&\qquad \qquad  x(1-e^{-\beta x})  \dv x .
\end{align*}
\end{lemma}
\subsection{Mean $\SINR$} 
We now present the mean $\SINR$ at the typical UE. We utilize the auxiliary variable trick by considering a RV $F$ with unit mean exponential distribution independent of $\SINR$. Hence, due to independence of these two RVs, we have 
\begin{align*}
		&\expect{\SINR}=\expect{F\cdot \SINR}\\
		&=\int \prob{F\cdot \SINR>\tau}\dv \tau=\int_{\tau=0}^{\infty}\prob{\frac{F {\mathcal{S}}(R_{\ob})}{I_{\brm}+I_{\srm}+N_{\ob}}>\tau}\dv \tau\\
		&=\int_{0}^{\infty}\expect{\expS{-\frac{\tau}{{\mathcal{S}}(R_{\ob})}\left(I_{\brm}+I_{\srm}+N_{\ob}\right)}}\dv \tau\\
		&=\int_{0}^{\infty}\expects{R_{\ob}}{\laplace_{I_{\rm b}\vert R_{\ob}}\left(\frac{\tau}{{\mathcal{S}}(R_{\ob})} \right)\laplace_{I_{\srm}}\left(\frac{\tau}{{\mathcal{S}}(R_{\ob})}\right)e^{-\frac{\tau N_{\ob}}{{\mathcal{S}}(R_{\ob})}}}\dv \tau.
	\end{align*}
Now utilizing the LT of BS and SE interferences, we get the mean SINR.

\subsection{Signal to SE Interference plus Noise Ratio (SSIR)}
	In order to highlight the impact of SE interference, we define ${\rm SSIR}={{\mathcal{S}}(R_{\ob})}/{(I_{\srm}+N_{\ob})},$ where  ${\mathcal{S}}(R_{\ob})$ is provided in \eqref{Sro} and $I_{\srm}$ is the SE interference. We derive the probability that SSIR at the typical UE is greater than a threshold i.e.
\begin{align}
	{\rm \tilde p_{c}}(\tau)=\prob{{\rm SSIR}>\tau}.\label{basiccoverageprobforSISR}
\end{align}
\begin{lemma}
	The SSIR CCDF $	{{\rm \tilde p_{c}}}(\tau)$ at the typical UE is given as $	{{\rm \tilde p_{c}}}(\tau)={1}/{2}-({1}/{\pi})	\int_{r=0}^{\infty}\int_{u=0}^{\infty}$
		\begin{align}
\mathrm{Im}\left[e^{-ju {\mathcal{S}}(r)+ju\tau N_{\ob}}\laplace_{I_{{\srm}}}(-ju{\tau})\right]({\dv u}/{u})f_{R_{\ob}}(r)\dv r. \label{eq:ssirCCDF}   
		\end{align}
\end{lemma}
\subsection{Rate CP and Average Ergodic  Rate}\label{sec:rate}
RC is defined as the probability that the instantaneous rate is above a certain threshold $\rho$. {The instantaneous rate of the typical UE is $B\log_{2}(1+\SINR)$, where $B$ denotes the bandwidth available to the typical UE. The mean load of the BS is $A_{\rmL}\lambda_{\urm}/\lambda_{\brm}$ where $A_{\rmL} \lambda_\urm$ denotes the density of associated users. Note that the serving BS of the typical UE carries a larger load on average, since a UE is more likely to lie in a larger cell. Hence, the mean load on the serving BS is $\overline{N}\approx1+1.28\,A_{\rmL}\lambda_{\urm}/\lambda_{\brm}$, where the factor $1.28$ accounts for the area bias \cite{singh2015joint}.}  Assuming that the total bandwidth $W$ is shared equally among the UEs served by a BS, the per-user bandwidth $B=W/\overline{N}$, 
\begin{theorem}\label{ratecoveragethm}
	The rate CP $\pt_{\mathrm{R}}(\rho)${,} the average ergodic rate $\overline{\mathcal{R}}$ and the area spectral efficiency ${\mathcal{A}}$ are respectively given as
	\begin{align*}
		\pt_{\mathrm{R}}(\rho)=\pt_{\crm}(2^{\rho/B}-1),
		\overline{\mathcal{R}}={\frac{B}{\ln 2}}\int_{0}^{\infty}\frac{\pt_{\crm}(\tau)}{\tau+1}\dv \tau, {\mathcal{A}}={\frac{\lambda_{\brm}\overline{\mathcal{R}}}{B}}.
	\end{align*}
\end{theorem}
\begin{proof}
	{The first expression follows from the definitions of  RC and  CP, since $B\log_{2}(1+\SINR)>\rho\implies$
$\SINR>2^{\rho/B}-1$. The second follows from the identity $\mathbb{E}[X]=\int_{0}^{\infty}\prob{X>x}\dv x$ applied to $X=\log_{2}(1+\SINR)$ and  substitution $\tau=2^{x}-1$. The third is  from the definition
 \cite{alouini1999area}.}
\end{proof}

\section{{Numerical Results}}\label{sec:numerical}
We now present numerical studies by evaluating the analytical expressions to derive insights for THz networks. 
Unless otherwise stated, all parameters take the {default} values listed in Table~\ref{tab:1}.
To validate the analysis and various approximations, we perform {extensive} Monte Carlo simulations  incorporating the correlation among scattering power across SEs. In each iteration, we generate two PPPs to model the spatial distribution of BSs and SEs. The blockage state of each link (BS--SE, SE--UE, or BS--UE) is determined using a Bernoulli RV with success probability $\losprob{x}{\rmL}=e^{-\beta x}$ where $x$ is the distance between the two points. SE orientations are randomly generated within the range $(0,\pi)$. For each SE, we calculate the incident angle from the BS based on the SE's orientation and the angular location of the BS relative to the SE. Similarly, the scattering angle is computed considering the SE's orientation and the angular location of the UE relative to the SE. The gain received by each SE and UE from each BS is allocated using the probabilities of the major and minor lobes of the BS antenna. Subsequently, the power incident on each SE from all BSs (on the same side of the SE as the UE)  is determined. 
 The scattered power from the SE is then distributed in all directions, and the interference power received by the typical UE is computed accordingly. The desired signal power is obtained from the nearest LOS BS. Next, we compute the interference power received by the UE from all LOS and NLOS BSs.  Finally, the SINR at the UE is calculated using the desired signal power, interference from BSs, SE interference, and noise. 

\inserttable

\begin{figure}[t!]
\centering
		\includegraphics[width=.45\linewidth,trim=20 0 0 0,clip]{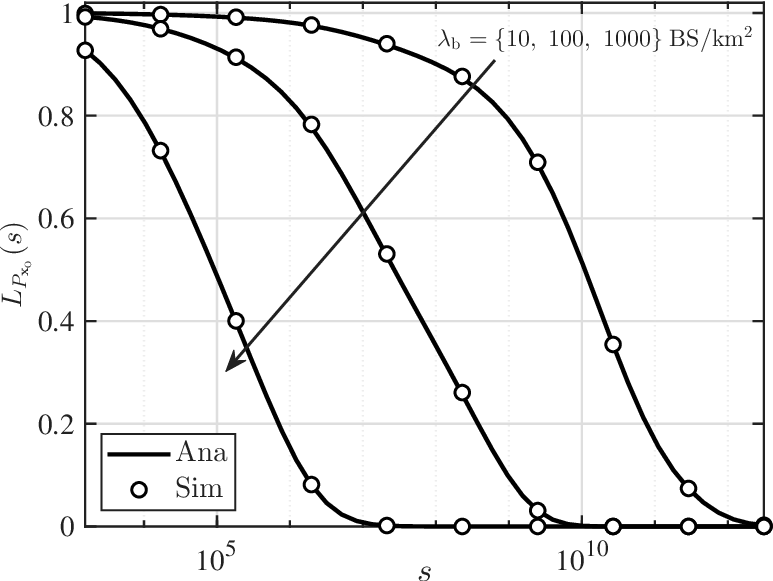}
		\includegraphics[width=.45\linewidth,trim=20 0 0 0,clip]{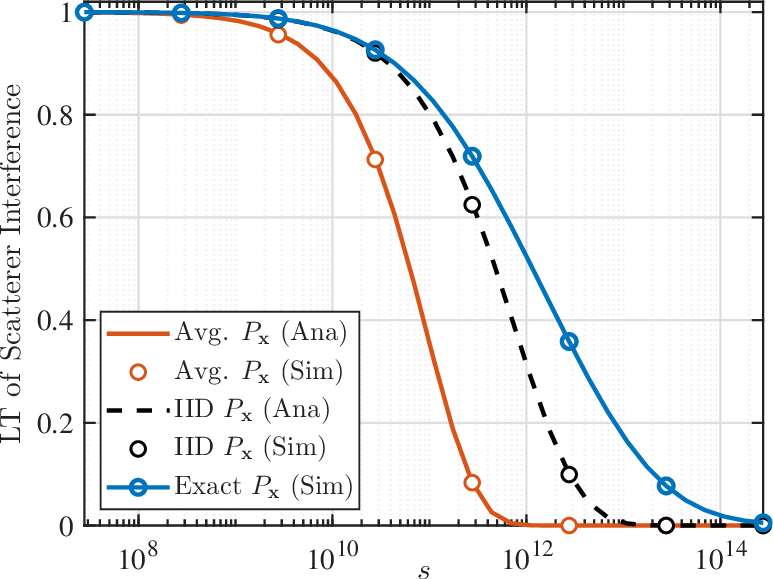}
	\vspace{-.12in}\\
	{\bf\footnotesize (a)}\hspace{1.4in}{\bf\footnotesize(b)}\vspace{-.1in}
	\caption{Analytical and simulation results for the LT of (a) the incident power {$P_{\x_\ob}$} on the typical SE {and} (b) the SE interference.}\label{fig:lt_and_interference1}\vspace{-.1in}
\end{figure}

\begin{figure}[t!]
	\ \ \includegraphics[width=0.45\linewidth,trim=20 0 0 0,clip]{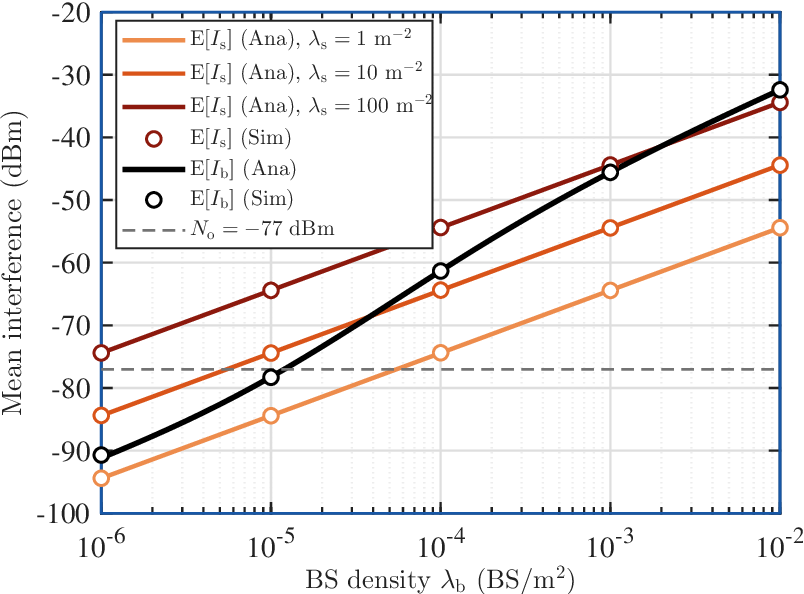}\hfill
	\includegraphics[width=0.45\linewidth,trim=20 0 0 0,clip]{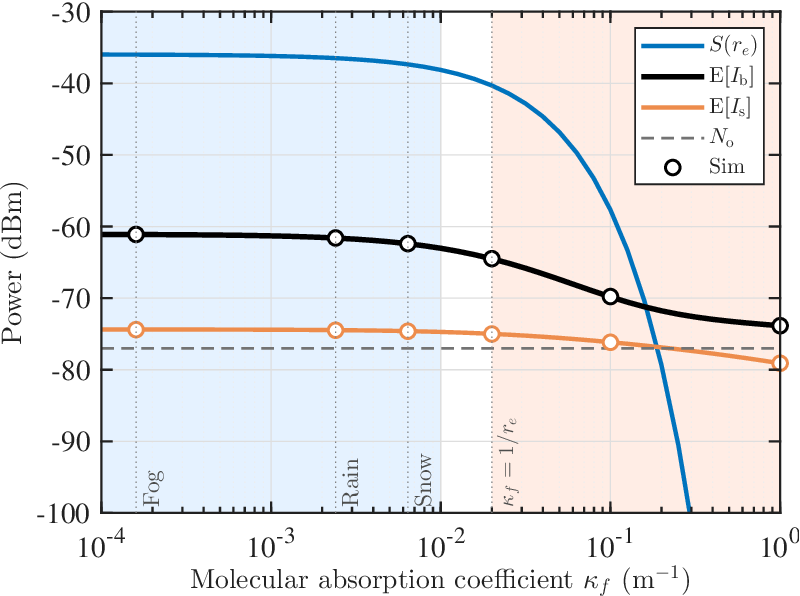}
\vspace{-.2in}\\
	{\bf\footnotesize (a)}\hspace{1.6in}{\bf\footnotesize(b)}\vspace{-.05in}
	
	\caption{(a) Mean SE interference $\expect{I_{\srm}}$ for $\lambda_{\srm}\in{\{1,10,100\}}$ \si{SE/m^2}, mean BS interference $\expect{I_{\brm}}$ and the noise $N_{\ob}$ against the BS density.
		(b) $\expect{I_{\srm}}$, $\expect{I_{\brm}}$, the received power $\mathcal{S}(r_{e})$ from the serving BS at the typical cell radius $r_{e}=1/(2\sqrt{\lambda_{\brm}})$ {versus} the molecular absorption coefficient $\kappa_{f}$.}
	\label{fig:combined_interference}
\vspace{-.2in}
\end{figure}
\subsectionu{{Distribution of Scattering Interference}}
In Fig. \ref{fig:lt_and_interference1}(a), we show the LT of the incident power ({from} Theorem~\ref{thm4}) at the typical SE for different BS densities {around $\lambda_{\brm}=\beta^{2}/4=100$~\si{BS/km^2}}. The analytical results closely match the Monte Carlo simulations. The LT decreases as $\lambda_{\brm}$ increases{, as} a denser network {causes more power to fall at} the typical SE. In Fig. \ref{fig:lt_and_interference1}(b), we show the LT of the SE interference {at the typical UE, along with the analytical values obtained via the two approximation approaches}. We can see that the IID $P_\x$ approximation gives more accurate values compared to the mean $P_\x$ approximation, as replacing $P_\x$ by its mean $m_{\ob}$ reduces the variance in the interference. Further, both approximations ignore the correlation and hence underestimate the spread of $I_{\srm}$. Both approximations preserve the mean $\expect{I_{\srm}}$ exactly as given by Theorem~\ref{theorem2}, which can be seen from the slope at $s\to0$.

\subsectionu{Impact of  BS Density and Atmosphere on $I_\srm$}
{Fig.~\ref{fig:combined_interference}(a) shows the variation of the mean scattering interference $\expect{I_{\srm}}$ with the BS density, along with the mean BS interference $\expect{I_{\brm}}$ and the noise floor $N_{\ob}$.} Since $\expect{I_{\srm}}$ is linear in $\lambda_{\brm}$ and $\lambda_{\srm}$ (Theorem~\ref{theorem2}), the three SE curves are straight lines {showing a linear increase with $\lambda_{\brm}$}. The BS interference $\expect{I_{\brm}}$ increases super-linearly {with $\lambda_{\brm}$ due to the dual path-loss effect \cite{andrews2023introduction}, where densification results in more links turning LOS}. The SE interference becomes significant relative to the noise in dense deployments. In Fig. \ref{fig:combined_interference}(b), we plot the mean {scattering} interference for three  weather conditions at $0.2$~\si{THz} \cite{moleculecoeff}:
\begin{enumerate}
	\item Fog condition with $100$ \si{m} visibility: 
  $\kappa_f= 0.00016$ {\si{m^{-1}}},
	\item Rain condition with  $20$ \si{mm/hr} rain: $\kappa_f=0.0024$ {\si{m^{-1}}},
	\item Snow at a rate of $100$ \si{mm/hr} with visibility of $770$  \si{m}, $\kappa_f=0.0064$ {\si{m^{-1}}}.
\end{enumerate}
We {also} plot the mean BS interference and the received power $\mathcal{S}(r_{e})$ from a serving BS at the typical cell radius $r_{e}=1/(2\sqrt{\lambda_{\brm}})${, to show how atmospheric absorption affects the serving link and the two interferences. Here we consider a BS density of $\lambda_{\brm}=\beta^{2}/4$ such that the typical cell radius becomes equal to} the LOS range $1/\beta=50$ \si{m}. {Such a density is usually preferred from the coverage perspective, as it corresponds to having only one LOS BS with the rest of the interfering BSs NLOS in an average sense \cite{andrews2023introduction}.} The absorption coefficient {affects} every link only through $e^{-\kappa_{f}z}${, with $z$ being the link length}.
{From Fig.~\ref{fig:combined_interference}(b) we can observe that for lower values of $\kappa_{f}$ (\ie $\kappa_{f}\ll 1/r_{e}$), the serving link and the dominant interfering links are not affected by molecular absorption. However, absorption starts to matter when $\kappa_{f}\gtrsim 1/r_{e}$. While it reduces all three of the serving signal, the scattering interference and the BS interference, the extent is significantly different. We can observe that the desired signal decreases fastest, followed by the BS interference, while the impact on the SE interference is the smallest.} At $\kappa_f=0.1$ \si{m^{-1}}, the desired signal, BS interference, and SE interference decrease by $22$, $11$, and $5$ \si{dB}, respectively. {Since} atmospheric absorption attenuates the desired signal more than the interference, {it} cannot be relied upon to suppress SE or BS interference.

\begin{figure}[t!]
	\centering
	\begin{minipage}[t]{0.5\linewidth}
		\includegraphics[width=\linewidth]{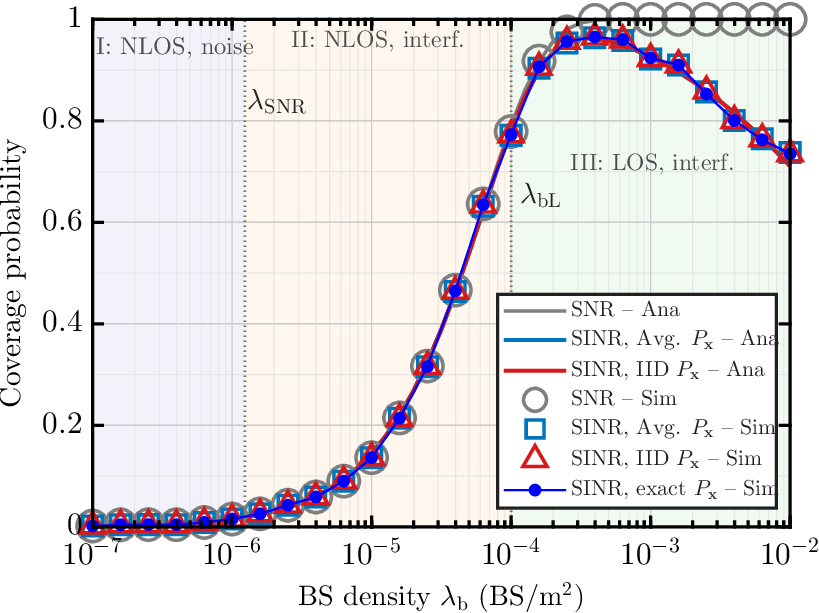}
\footnotesize{		 (a)\! $\beta=0.02$~m$^{-1}$:
		$\lambda_{\rm SNR}\!<\!\lambda_{\rm bL}$}
	\end{minipage}\hfill
	\begin{minipage}[t]{0.5\linewidth}
		
		\includegraphics[width=\linewidth]{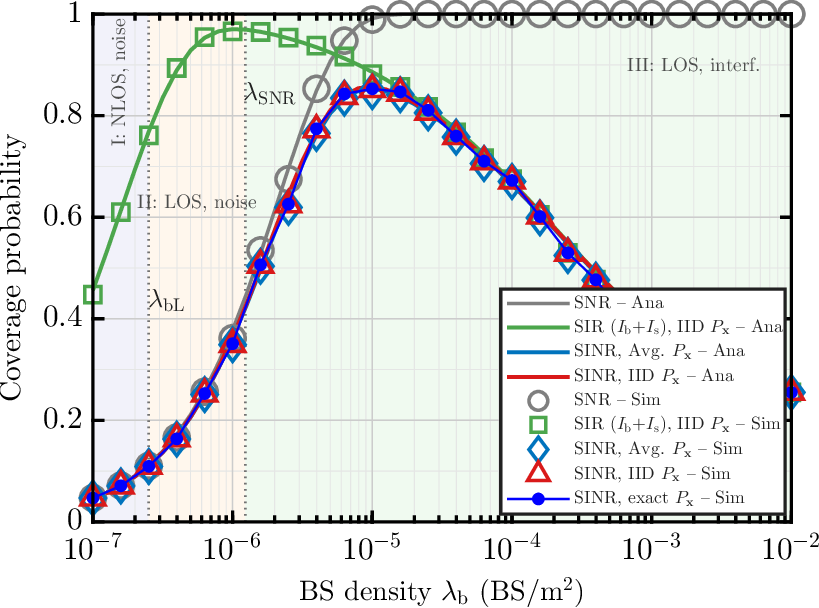}
\footnotesize{(b)\! $\beta=0.001$~m$^{-1}$:
		$\lambda_{\rm bL}\!<\!\lambda_{\rm SNR}$}
	\end{minipage}
	\caption{CP against the BS density $\lambda_{\rm b}$ at $\tau=20$~\si{dB} for (a) $\beta=0.02$~m$^{-1}$ and (b) $\beta={0.001}$~m$^{-1}$. Two {critical density values}, $\lambda_{\rm bL}$ and $\lambda_{\rm SNR}$, shape each curve and the densification regions. At $\lambda_{\rm SNR}$, the serving link overcomes the noise, and at $\lambda_{\rm bL}$, the interfering BSs start to become LOS.}\label{CPandmeanSINR}
	\vspace{-.15in}
\end{figure}
\subsectionu{Impact of Densification {and Densification Regimes}}
Fig.~\ref{CPandmeanSINR} shows  CP versus BS density $\lambda_{\brm}$ at $\tau=20$~\si{dB} for two blockage environments. Fig.~\ref{CPandmeanSINR}(a) considers a densely blocked environment with $\beta=0.02$~\si{m^{-1}} and LOS range $1/\beta=50$~\si{m}, while Fig.~\ref{CPandmeanSINR}(b) considers an open environment with {$\beta=0.001$~\si{m^{-1}} and LOS range $1/\beta=1000$~\si{m}}. Both figures also show SNR and SINR coverage for the two $P_{\x}$ approximations. Fig.~\ref{CPandmeanSINR}(b) also shows SIR coverage to isolate the effect of interference. {Let us define $\lambda_{\rm SNR}=1/(4r_{\rm cov}^{2})$ with $r_{\rm cov}$ being the distance at which the SNR meets the threshold $\tau$, \ie $\mathcal{S}(r_{\rm cov})=\tau N_{\ob}$ in~\eqref{Sro}. Hence, $\lambda_{\rm SNR}$ denotes the density above which the average serving link is short enough for the received signal to overcome the noise, and $\lambda_{\brm}<\lambda_{\rm SNR}$ represents the noise-limited regime. Further, let us define $\lambda_{\rm bL}=\beta^{2}/4$, such that the mean typical cell radius equals the LOS range, \ie $r_{e}=1/\beta$. Hence, for $\lambda_{\brm}\ll\lambda_{\rm bL}$ all links are NLOS, while for $\lambda_{\brm}\gg\lambda_{\rm bL}$ all links, including the serving link, are usually LOS in a probabilistic sense. Around $\lambda_{\brm}\approx\lambda_{\rm bL}$, the serving link is usually LOS while the rest of the interfering links are most probably NLOS. Without noise, the coverage peaks slightly above $\lambda_{\rm bL}$, where a LOS serving BS is almost always available, and it decreases when the BS density is increased or decreased from there \cite{andrews2023introduction}. Now, as we densify the network, the effective coverage is due to the combined effect of both factors, the SNR trend and the SINR trend, as characterized by the two critical density values $\lambda_{\rm SNR}$ and $\lambda_{\rm bL}$.} {Note that $\lambda_{\rm SNR}=1.2$~\si{BS/km^2}, while $\lambda_{\rm bL}=100$~\si{BS/km^2} in the dense environment and} {$0.25$~\si{BS/km^2} in the open environment. Hence $\lambda_{\rm SNR}<\lambda_{\rm bL}$ in Fig.~\ref{CPandmeanSINR}(a) and $\lambda_{\rm bL}<\lambda_{\rm SNR}$ in Fig.~\ref{CPandmeanSINR}(b). In both panels, the coverage passes through the three regions marked as I, II and III, with the middle region different between two panels.} In region I, the serving link cannot overcome the noise even without interference and the CP is low. {In Fig.~\ref{CPandmeanSINR}(a), the serving signal clears the noise in region II while the interfering BSs are still mostly NLOS. Hence the CP rises steeply with densification, as more UEs find a LOS serving BS. In region III, the interfering BSs also turn LOS and their aggregate power grows faster than the serving signal. Hence the CP falls beyond an optimal density located slightly above $\lambda_{\rm bL}$. In Fig.~\ref{CPandmeanSINR}(b), the interfering BSs turn LOS in region II, while the serving link is still limited by the noise. However, the SIR coverage still rises in this region, since more UEs find a LOS serving BS, and it peaks near $\lambda_{\rm SNR}$. In region III, the SIR coverage falls while the SNR coverage is still rising, and the CP peaks above $\lambda_{\rm SNR}$. The peak is lower than in Fig.~\ref{CPandmeanSINR}(a), since the interfering BSs are LOS before the serving link overcomes the noise. The optimal density is therefore set by the blockage in the dense environment and by the link budget in the open environment.} Beyond the optimum, the fall is steeper in the open environment, where nearly every interferer within the larger LOS range becomes visible to the UE, and milder in the dense environment, where the blockages keep shielding most of the interferers. {Further, the SNR curve rises monotonically and saturates, while the SINR curves separate from it beyond the larger of the two critical densities. Hence the interference affects the coverage only beyond this density. In Fig.~\ref{CPandmeanSINR}(b), the SIR curve lies well above the SINR curves in region II, since the noise still limits most UEs there. It merges with them beyond the optimum, where only the interference sets the coverage.} {Both Figs.~\ref{CPandmeanSINR}(a) and \ref{CPandmeanSINR}(b) thus identify different densification regimes. In particular, the middle region differs between them. Since both densities are fixed by the system parameters, their order identifies the regime of a deployment before any coverage curve is computed.} Across both panels the mean and the IID $P_{\x}$ approximations are indistinguishable from the exact simulation, {which validates the approximations considered in the paper.} 

\begin{figure}
\centering
	\includegraphics[width=0.48\linewidth,trim=15 0 0 0,clip]{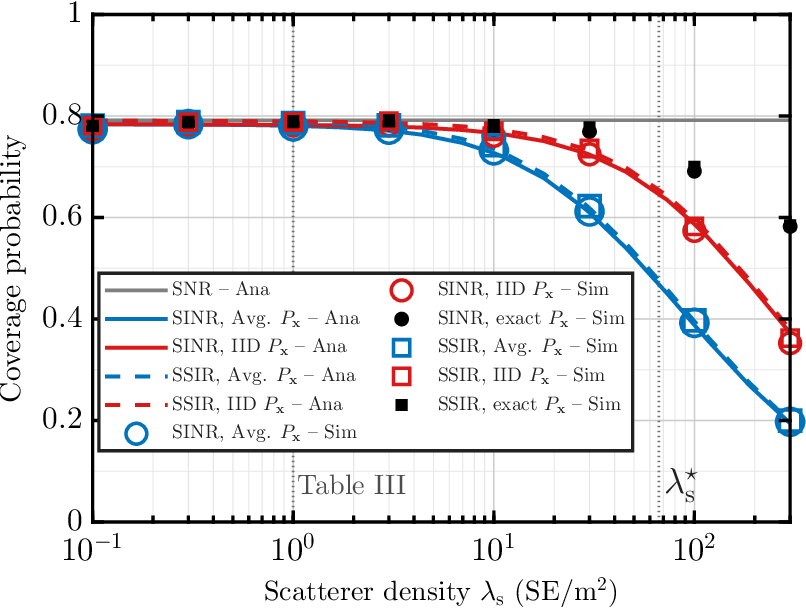}\ \ 
		\includegraphics[width=0.46\linewidth,trim=15 0 0 0,clip]{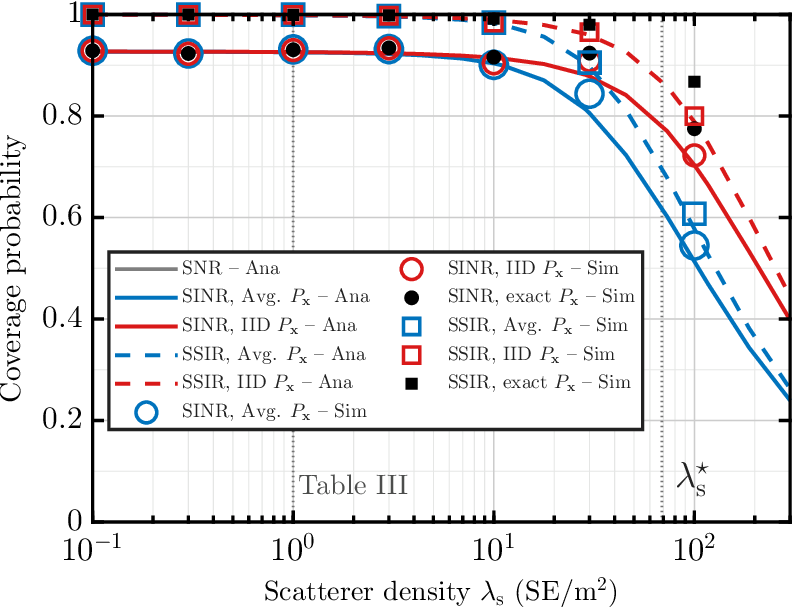}
	\vspace{-.2in}\\
	{\bf\footnotesize (a)}\hspace{1.5in}{\bf\footnotesize(b)}\hfill\vspace{-.05in}
	
	\caption{{CP ${\rm p_{c}}$ (solid) and SSIR CCDF ${\rm \tilde p_{c}}$ (dashed) against the SE density $\lambda_{\srm}$ at $\tau=20$~\si{dB} for (a) $\lambda_{\brm}=10^{-4}$ and (b) $10^{-3}$~\si{BS/m^2}. The  critical density $\lambda_{\srm}^{\star}$ is the density at which $\tau\expect{I_{\srm}}=\mathcal{S}(r_{e})$.}}
	\label{fig:fig8}\vspace{-.1in}
\end{figure}
\begin{figure}[t!]
	\centering
	\includegraphics[width=0.52\linewidth]{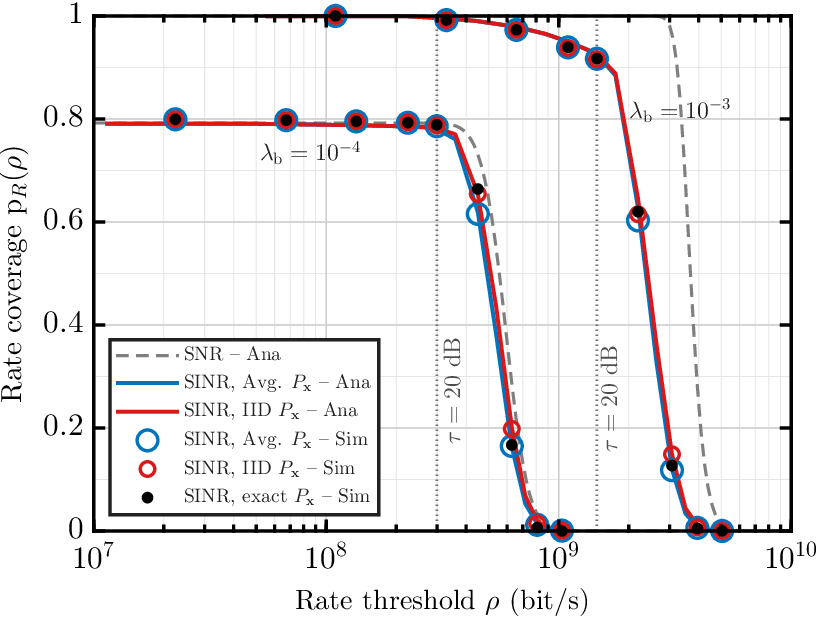}\vspace{-.05in}
	\caption{{RC $\pt_{\mathrm{R}}(\rho)$ against the rate threshold $\rho$ at $\lambda_{\brm}=10^{-4}$ and $10^{-3}$~\si{BS/m^2}, with the per user bandwidth $B=W/\overline{N}$.}}
	\label{fig:ratecov}\vspace{-.15in}
\end{figure}

\subsectionu{Impact of SE Density}
{Fig.~\ref{fig:fig8} isolates the effect of scattering by varying the SE density $\lambda_{\srm}$ at the two fixed BS densities $\lambda_{\brm}=10^{-4}$~\si{BS/m^2} in (a) and $10^{-3}$~\si{BS/m^2} in (b), to show the variation due to the SE interference only. In (a) the SINR and the SSIR curves nearly coincide, since at this density the outage is dominated by the absence of a LOS serving BS and the BS interference makes a negligible contribution to the outage. In Fig.~\ref{fig:fig8}(b), the BS interference contributes significantly to the outage, and hence, the SINR curve stays below the SSIR curve over the whole range. We define the critical density $\lambda_{\srm}^{\star}$ by $\tau\expect{I_{\srm}}=\mathcal{S}(r_{e})$, \ie the density at which the ratio of the signal at the typical cell edge $r_{e}=1/(2\sqrt{\lambda_{\brm}})$ to the mean SE interference equals $\tau$, which can be computed in closed form using \eqref{Sro} and Theorem~\ref{theorem2}. For the Table~\ref{tab:1} parameters, $\lambda_{\srm}^{\star}\approx67$~\si{SE/m^2}. Note that since both $\expect{I_{\srm}}$ and $\mathcal{S}(r_{e})$ grow linearly with $\lambda_{\brm}$, $\lambda_{\srm}^{\star}$ is nearly independent of $\lambda_{\brm}$. We can observe that the coverage starts to fall when SE density is comparable to $\lambda_{\srm}^{\star}$. This regime corresponds to scatterers covering a fraction $\lambda_{\srm}^{\star}A_{\srm}\approx0.7$ of the area. This shows that in heavily cluttered environments, scattering interference must be included in the analysis. At very large $\lambda_{\srm}$, the exact $P_{\x}$ simulation lies above both approximations, so the analysis is conservative there.}
\subsectionu{{Rate Coverage}}
{Fig.~\ref{fig:ratecov} shows the {variation of the} RC $\pt_{\mathrm{R}}(\rho)$ with the rate threshold $\rho$.} {The two sets of curves correspond to $\lambda_{\brm}=10^{-4}$ and $10^{-3}$~\si{BS/m^2}, with per-user bandwidths $B=44.9$ and $219.3$~\si{MHz}. For small $\rho$, the RC is constant and equals the probability that the UE is served, which is set by the LOS availability. As $\rho$ increases, the RC falls at a value of $\rho$ set by the per user bandwidth, since the rate $B\log_{2}(1+\SINR)$ is proportional to $B$ but varies slowly with the $\SINR$. Hence, densification improves the rate mainly by increasing the bandwidth available to each user.}

\section{Conclusion}
In this paper, we analyzed the coverage of a THz network in the presence of blockages, SEs, directional antennas and molecular absorption. We modeled the BSs and the SEs as PPPs and the blockages as a Boolean model, and we included the antenna side lobes and the random orientation of the SEs. We derived the LT of the interference from the SEs and from the BSs, and from these we obtained the CP, the mean SINR, the RC and the area spectral efficiency. We validated the analysis against a Monte Carlo simulation that retains the correlation of the incident power across SEs. The results reduce the network design to a few quantities that follow from the system parameters. {We identified the densification regimes of the network through two critical BS densities, and showed that the shape of the coverage versus densification curve and the optimal BS density depend on these two densities.} {A critical SE density, obtained in closed form, marks where scattering must be accounted for. It is nearly independent of the BS density and corresponds to the scatterers covering a substantial fraction of the area, so scattering must be modeled in heavily cluttered indoor deployments. The RC shows that the achievable rate is limited by the availability of a LOS link and by the per user bandwidth rather than by the SINR, so densification improves the rate through the bandwidth reuse.}

 \appendix 


\subsection{Proof of Theorem \ref{thm:ms}}\label{app:thm:ms}
\vspace{-.2in}
\begin{align*}
	\textstyle\mathbb{E}\left[{P}_{\x_\ob}\right] &=\mathbb{E}\left[\sum\nolimits_{{\z_{i}}\in \Phi_{\rmL,\brm}^+}P_{\trm}g_{\trm}(\widetilde{\beta}_{\z_{i}})\ell_{\rmL,\srm}(\|{\z_{i}}\|)\cos(\theta_{\z_i})\right.\\
	&\textstyle\left.+\sum\nolimits_{{\z_{i}}\in \Phi_{\rmN,\brm}^+}P_{\trm}g_{\trm}(\widetilde{\beta}_{\z_{i}})\ell_{\rmN,\srm}(\|\z_{i}\|)\cos(\theta_{\z_i})\right].
\end{align*}Let $\tilde{\Phi}=\left\{\left({\z}_i, \phi_{\z_i}\right): {\z}_i \in \Phi_{\brm}\right\}$ be a marked PPP with underlying PP $\Phi_{\brm}$ being a PPP. From \eqref{tildebetapibypi} replacing $\widetilde{\beta}_{\z_i}$, we get  
\begin{align*}
	\textstyle\mathbb{E}\left[{P}_{\x_\ob}\right] &=\mathbb{E}\left[\sum\nolimits_{\z_{i}\in \Phi_{\rmL,\brm}^+}P_{\trm}g_{\trm}(\widetilde{\beta}(\delta_{\z_i},\phi_{\z_i}))\ell_{\rmL,\srm}(\|\z_{i}\|)\cos(\theta_{\z_i})\right.\\
	&\textstyle\left.+\sum\nolimits_{\z_{i}\in \Phi_{\rmN,\brm}^+}P_{\trm}g_{\trm}(\widetilde{\beta}(\delta_{\z_i},\phi_{\z_i}))\ell_{\rmN,\srm}(\|\z_{i}\|)\cos(\theta_{\z_i})\right].
\end{align*}Here, $\phi_{\z_i}$ is the mark of the corresponding PP. We derive the average of incident power due to LOS BSs using Campbell's theorem given as
\begin{align*}
	&\textstyle\expect{P_{\x_\ob,\rmL}}=({P_{\trm}}/{(2\pi)})\int_{\theta_t=-\pi/2}^{\pi/2}\int_{\phi_{\trm}=0}^{2\pi}\int_{0}^{\infty}\lambda_{\rmL,\brm}(z)\ell_{\rmL,\srm}(z)\\
	&\textstyle\times g_{\trm}(\widetilde{\beta}({\delta},{\phi_{\trm}})) z\dv z {\dv \phi_{\trm}}\cos(\theta_t)\dv \theta_t\\
	&\textstyle\stackrel{(a)}=({P_{\trm}}/{(2\pi)})\int_{\delta=0}^{\pi/2}\!\int_{0}^{\infty}\! \lambda_{\rmL,\brm}(z)\ell_{\rmL,\srm}(z)\left(\int_{\phi_{\trm}=0}^{2\pi}g_{\trm}(\widetilde{\beta}(\delta,\phi_{\trm})){\dv \phi_{\trm}}\right)\\
	&\textstyle\times z\dv z \sin(\delta)\dv \delta+({P_{\trm}}/{(2\pi)})\int_{\delta=\pi/2}^{\pi}\int_{0}^{\infty}\lambda_{\rmL,\brm}(z)\ell_{\rmL,\srm}(z)\times\\
	&\textstyle\left(\int_{\phi_{\trm}=0}^{2\pi}g_{\trm}(\widetilde{\beta}(\delta,\phi_{\trm})){\dv \phi_{\trm}}\right)  z\dv z \sin(\delta)\dv \delta,
\end{align*}

\noindent where step $(a)$ is obtained by replacing $\theta_t$. 
In the next step, we solve the integral for $g_{\trm}(\widetilde{\beta}(\delta,\phi_{\trm}))$ given as  
\begin{align*}
	\textstyle\int_{\phi_{\trm}=0}^{2\pi}g_{\trm}(\widetilde{\beta}(\delta,\phi_{\trm}))\dv \phi_{\trm}&\textstyle=\int_{\phi_{\trm}=0}^{2\pi}g_{\trm}(\pi-\delta+\phi_{\trm})\dv \phi_{\trm}.
\end{align*}
From \eqref{Gain}, for $\vert{\widetilde{\beta}(\delta,\phi_{\trm})}\vert\leq \gamma_{\rm t}/2, g_{\trm}(\widetilde{\beta}(\delta,\phi_{\trm})) =\Grm_{\mrm,\trm}$. Hence, for $-\gamma_{\trm}/2\leq(\pi-\delta+ \phi_{\trm})\leq \gamma_{\trm}/2,\, g_{\trm}(\widetilde{\beta}(\delta,\phi_{\trm})) =\Grm_{\mrm,\trm}$ and otherwise $\Grm_{\srm,\trm}$.
\begin{align}
	&\textstyle\int_{\phi_{\trm}=0}^{2\pi}g_{\trm}(\widetilde{\beta}(\delta,\phi_{\trm}))\dv \phi_{\trm}=\int_{\phi_{\trm}=-\pi+\delta-\gamma_{\trm}/2}^{-\pi+\delta+\gamma_{\trm}/2}\Grm_{\mrm,\trm}\dv \phi_{\trm}\nonumber\\
	&\textstyle+\int_{\phi_{\trm}=0}^{-\pi+\delta-\gamma_{\trm}/2}\Grm_{\srm,\trm}\dv \phi_{\trm}+\int_{\phi_{\trm}=-\pi+\delta+\gamma_{\trm}/2}^{2\pi}\Grm_{\srm,\trm}\dv \phi_{\trm}\nonumber\\
	&\textstyle=\Grm_{\mrm,\trm}\gamma_{\trm}+\Grm_{\srm,\trm}(2\pi-\gamma_{\trm}).
\end{align}Similarly, we can check for other cases of $\widetilde{\beta}(\delta,\phi_{\trm})$. Therefore. the average incident power on the typical SE from LOS BSs is
\begin{align*}
	&\textstyle\expect{P_{\x_\ob,\rmL}}=\!\frac{P_{\trm}\left(\Grm_{\mrm,\trm}\gamma_{\trm}+\Grm_{\srm,\trm}(2\pi-\gamma_{\trm})\right)}{2\pi}\int_{0}^{\infty}\lambda_{\rmL,\brm}(z)\ell_{\rmL,\srm}(z) z \dv z\\
	&\textstyle+\frac{P_{\trm}\left(\Grm_{\mrm,\trm}\gamma_{\trm}+\Grm_{\srm,\trm}(2\pi-\gamma_{\trm})\right)}{2\pi}\int_{0}^{\infty} \lambda_{\rmL,\brm}(z)\ell_{\rmL,\srm}(z) z \dv z.
\end{align*}Similarly we find the average value of $P_{\x_\ob,\rmN}$.

\subsection{Proof of Theorem \ref{theorem 4} }\label{app:D}
The LT of  $I_{\srm,\app1}$ is $\laplace_{I_{\srm,\app1}}(s)=\expect{\expS{-sI_{\srm,\app1}}}$ 
\begin{align*}
	&\textstyle=\expect{\prod\nolimits_{\x_{i}\in \Phi_{\srm}}\!\!\!\expS{-s S^{2}m_{\ob}g_{\rrm}(\widetilde{\beta}_{\x_i})\ell_{\lstate{\x_i\ueb},\ue}(\|\x_{i}\|)\sprof{\cos\psi_{\x_i}}}}\\
	&\textstyle=\prod_{\lstate{}\in\{\rmL,\rmN\}}\expect{\prod_{\x_{i}\in \Phi_{\lstate{},\srm}}
		\expU{\!-s S^{2}m_{\ob}g_{\rrm}(\widetilde{\beta}_{\x_i})
			\pathlossSEtoUs{\lstate{}}
			(\|\x_{i}\|)\sprof{\cos\psi_{\x_i}}}\!}.
\end{align*}
The first term represents LOS SEs given as $\laplace_{I_{\srm,\rmL,\app1}}(s)=$
\begin{align*}
	&\textstyle\expect{\prod\nolimits_{\x_{i}\in \Phi_{\rmL,\srm}}\expS{-s S^{2}m_{\ob}g_{\rrm}(\widetilde{\beta}_{\x_i})\ell_{\rmL,\ue}(\|\x_{i}\|)
			\sprof{\cos(\psi_{\x_i})}}}\\
	&\textstyle\stackrel{(a)}=\exp\left(-\lambda_{\srm}\int_{x=0}^{\infty}\int_{\Omega=0}^{2\pi}\left(1\right.\right.\\
	&\textstyle\left.\left.-\mathbb{E}_{\omega_\x}\left[e^{-s S^{2}m_{\ob}g_{\rrm}(\widetilde{\beta}_{\x})\ell_{\rmL,\ue}(x)
		\sprof{\cos\psi_\x}
	}\right]\right)e^{-\beta x}x\dv x\dv \Omega\right),
\end{align*}
where $(a)$ is due to  the PGFL of $\Phi_{\rmL,\srm}$.
Now, from  \eqref{psiequation} and Proposition \ref{propos2}, 	we have
\begin{align*}
	&\textstyle\laplace_{I_{\srm,\rmL,\app1}}(s)=\exp\left(-{\lambda_{\srm}}
	\int_{x=0}^{\infty}\int_{\Omega=0}^{2\pi}\int_{y=0}^{1}\right.\\
	&\textstyle\left.\left(1-e^{-s S^{2}m_{\ob}g_{\rrm}(\widetilde{\beta}_{\x})\ell_{\rmL,\ue}(x)\sprof{y}}\right)e^{-\beta x}\PDFY(y)x\dv y \dv x\dv \Omega\right).
\end{align*}
Now, noticing that $\widetilde{\beta}_{\x}=\phi_{\ueb}-\Omega_\x=-\Omega_\x$,  we get
\begin{align*}
	&\textstyle\laplace_{I_{\srm,\rmL,\app1}}(s)=\exp\left(-\sum\nolimits_{i=1}^{2}\pt_{\rrm,i}{\lambda_{\srm}}\int_{x=0}^{\infty}\int_{\Omega=0}^{2\pi}\int_{y=0}^{1}\right.\\
	&\textstyle\left.\left(1-e^{-s S^{2}m_{\ob}\Grm_{\rrm,i}
		\pathlossSEtoUs{\rmL}
		(x)
		\sprof{y}
	}\right)e^{-\beta x}
	\PDFY(y)x\dv y \dv x\dv \Omega\right).
\end{align*}The proof for the NLOS SEs is similar. 
\subsection{Relation between $\psi_\x$ and $\Omega_\x$}\label{app:psi_xproof}

{The scattering angle $\psi_\x$ may take different values depending on the angular position ($\Omega_\x$) and random orientation ($\omega_\x$) of the SE. To derive $\psi_\x$, we have shown a scenario in Fig. \ref{fig:scattererscenrio2}(c) for the case when $\Omega_\x<\pi/2$ and $\Omega_\x<\omega_\x$. In such case $\psi_\x={\pi}/{2}-(\omega_\x-\Omega_\x).$ Similarly, we can derive the $\psi_\x$ for other cases as well which is systematically provided below}
\begin{align}
\!\!	&\psi_\x\!\!=\!\!\begin{cases}
		{\pi}/{2}-(\Omega_\x-\omega_\x)&\!\!\!\! \!\!\text{if } \,  \Omega_\x>\omega_\x\,\text{and}\,0<\Omega_\x<\pi\\
		{\pi}/{2}-(\omega_\x-\Omega_\x)&\!\!\!\! \!\!\text{if } \,  \Omega_\x<\omega_\x\,\text{and}\,0<\Omega_\x<\pi\\
		{3\pi}/{2}-(\Omega_\x-\omega_\x)& \!\!\!\!\!\!\text{if } \, (\Omega_\x-\pi)>\omega_\x \,\text{and}\,\pi<\Omega_\x\!<2\!\pi\\
		-{\pi}/{2}+(\Omega_\x-\omega_\x)& \!\!\!\!\!\!\text{if } \, (\Omega_\x-\pi)<\omega_\x \,\text{and}\, \pi<\!\Omega_\x<\!2\pi.
	\end{cases}\nonumber
\end{align}

\subsection{Proof of Theorem \ref{LT_of_Ib}}\label{proof_of_LT_of_Ib}
The LT of $I_{\brm,\rmL}$ is $\laplace_{I_{\brm,\rmL}\vert R_\ob}(s)=\expect{\expS{-s I_{\brm,\rmL}}\vert R_\ob}$
\begin{align*}
	&\textstyle=\expect{\expS{-s \sum\nolimits_{\z_{i}\in \Phi_{\rmL,\brm}}P_{\trm}g_{\z_{i}}\pathlossBStoUs{\rmL}(\|\z_{i}\|)}}\\
	&\textstyle=\expect{\prod\nolimits_{\z_{i}\in \Phi_{\rmL,\brm}}\expS{-sP_{\trm}g_{\z_{i}}\pathlossBStoUs{\rmL}(\|\z_{i}\|)}}\\
	&\textstyle=\expects{\Phi_{\rmL,\brm}}{\prod\nolimits_{\z_{i}\in \Phi_{\rmL,\brm}}\expects{g_{\z_{i}}\vert\Phi_{\rmL,\brm}}{\expS{-sP_{\trm}g_{\z_{i}}\pathlossBStoUs{\rmL}(\|\z_{i}\|)}}}{.}
\end{align*}
Conditioned on $R_\ob$, the LOS interferers satisfy $\|\z_i\|>R_\ob$. Averaging over $g_{\z_i}$ using \eqref{eq:AntennaGainValues} and applying the PGFL of $\Phi_{\rmL,\brm}$ over $\|\z\|>R_\ob$, we get the LT of interference due to LOS BSs. Similarly, we can derive the LT of interference due to NLOS BSs. Finally, taking the product of the LTs, we get the LT of interference due to all the BSs.

\subsection{Proof of Corollary~\ref{Cor:LTIsIID}}\label{app:corIID}
\noindent{Substituting $s\!=\!-ju$ in the LOS component of Theorem~\ref{LTinterferecenapp2} 
}
\begin{align}
	&\textstyle\laplace_{I_{\srm,\rmL,\app2}}(-ju)=\exp\left(-\sum\nolimits_{i=1}^{2}{\pt_{\rrm,i}2\pi\lambda_{\srm}}\int_{0}^{\infty}\int_{0}^{1}\left(1- \right.\nonumber\right.\\
	&\textstyle\left.\left.\vphantom{\frac{num}{den}}\laplace_{P_{\x_\ob}}\left(-juS^{2}{\Grm_{\rrm,i}}
	\pathlossSEtoUs{\rmL}
	(x)
	\sprof{y}
	\right)\right)\PDFY(y)xe^{-\beta x}\dv y   \dv x\right)\label{eq:IsCorr52a}.
\end{align}
From Theorem~\ref{thm4}, $\laplace_{P_{\x_\ob}}\!(-ju)\!=\!\laplace_{P_{\x_\ob,\rmL}}\!(-ju)\laplace_{P_{\x_\ob,\rmN}}\!(-ju)$ with
\begin{align*}
	&\textstyle\laplace_{P_{\x_\ob,\lstate{}}}(-ju)=
	\exp\left(-\lambda_{\brm}\int_{0}^{\infty}
	\int_{0}^{\pi}\sum\nolimits_{i=1}^{2}
	\pt_{\trm,i}\right.\\
	&\textstyle\qquad\left.\left(1-{\expS{juP_{\trm}
	\Grm_{\trm,i}
	\pathlossBStoSEs{\lstate{}}
	(z)
	\sin (\delta)}}\right)
	\losprob{z}{\lstate{}}
	z \dv z \dv \delta\vphantom{\int_{0}^{\infty}}\right).
\end{align*}{Writing $e^{jt}=\cos t+j\sin t$ and collecting the real and imaginary terms gives $\laplace_{P_{\x_\ob,\lstate{}}}(-ju)=R_{\x_\ob,\lstate{}}(u)\expS{j \Theta_{\x_\ob,\lstate{}}(u)}$ with $R_{\x_\ob,\lstate{}}(u)$ and $\Theta_{\x_\ob,\lstate{}}(u)$ as in \eqref{rsl} and \eqref{thetasl}, and hence}
\begin{align}
	&\laplace_{P_{\x_\ob}}(-ju)=R_{\x_\ob}(u)\expS{j\Theta_{\x_\ob}(u)}.
	\label{eq:LPxju}
\end{align}{Substituting \eqref{eq:LPxju} in \eqref{eq:IsCorr52a} and again separating the real and imaginary parts of $R_{\x_\ob}(\cdot)e^{j\Theta_{\x_\ob}(\cdot)}$ gives $\laplace_{I_{\srm,\rmL,\app2}}(-ju)=R_{\srm,\rmL,\app2}(u)\expS{j \Theta_{\srm,\rmL,\app2}(u)}$ with $R_{\srm,\rmL,\app2}(u)$ and $\Theta_{\srm,\rmL,\app2}(u)$ as given in \eqref{RIsLapp2} and \eqref{TIsLapp2}. Applying the same steps to the NLOS component completes the proof.}

\subsection{Proof of Theorem \ref{coverage_probablity_general}}
\label{proof_of_coverage_probablity_general}
From the definition, we have $	{\rm p_{c}}(\tau)$
	\begin{align*}
	&=\prob{\frac{{\mathcal{S}}(R_{\ob})}{I_{\brm}+I_{\srm}+N_{\ob}}>\tau}=\prob{{{\mathcal{S}}(R_{\ob})}>({I_{\brm}+I_{\srm}+N_{\ob}})\tau}.%
	\end{align*}
	Using the Gil-Pelaez inversion theorem \cite{gil1951note} and averaging over
 $R_{\ob}$, the CP can be obtained as 
	\begin{align}\label{pctau}
		\pcov (\tau)&= \mathbb{E}_{R_{\ob}} \left[\frac{1}{2} - \frac{1}{\pi} \int_{u=0}^{\infty} {\mathrm{Im}\left[\mathbb{E}\left[e^{-j u {\mathcal{S}}^{\prime}}\vert R_{\ob}\right] e^{ju \tau N_{\ob}} \right]}\frac{\dv u}{u}\right], 		\nonumber
	\end{align}where ${\mathcal{S}}^{\prime} ={{\mathcal{S}}(R_{\ob}) - \tau (I_{\brm}+I_{\srm})}$. Note that
	\begin{align*}
		\mathbb{E} \left[e^{-ju{\mathcal{S}}^{\prime}}\vert R_{\ob} \right]= \mathbb{E}\left[ e^{-ju {\mathcal{S}}\left(R_{\ob}\right)}e^{ju\tau (I_{\brm}+I_{\srm})}\vert R_{\ob} \right].	
	\end{align*}
Applying the definition of the LT, we get the desired result.

\bibliographystyle{IEEEtran}
\bibliography{thz_refs}

\end{document}